\documentclass[final,5p,times]{elsarticle}
\usepackage{algorithm}
\usepackage{algpseudocode} 

\usepackage{tabularx, array}
\usepackage{graphicx}
\usepackage[T1]{fontenc}
\usepackage{amsmath,amsfonts}
\usepackage{multirow}
\usepackage[export]{adjustbox}
\usepackage{setspace}
\usepackage{subcaption} 
\usepackage{booktabs}
\usepackage{float}

\begin{document}
\begin{frontmatter}

\title{Integrating Wide and Deep Neural Networks with Squeeze-and-Excitation Blocks for Multi-Target Property Prediction in Additively Manufactured Fiber Reinforced Composites}
\author[1]{Behzad Parvaresh}
\ead{bparvaresh@smu.edu}
\cortext[cor1]{Corresponding author}

\author[1]{Rahmat K. Adesunkanmi}
\ead{radesunkanmi@smu.edu}

\author[1]{Adel Alaeddini\corref{cor1}}
\ead{aalaeddini@smu.edu}

\affiliation[1]{organization={Department of Mechanical Engineering, Southern Methodist University}, addressline={Dallas, Texas}, country={USA}}

\begin{abstract}
Continuous fiber-reinforced composite manufactured by additive manufacturing (CFRC-AM) offers opportunities for printing lightweight materials with high specific strength. However, their performance is sensitive to the interaction of process and material parameters, making exhaustive experimental testing impractical.
In this study, we introduce a data-efficient, multi-input, multi-target learning approach that integrates Latin Hypercube Sampling (LHS)-guided experimentation with a squeeze-and-excitation wide and deep neural network (SE-WDNN) to jointly predict multiple mechanical and manufacturing properties of CFRC-AMs based on different manufacturing parameters. We printed and tested $155$ specimens selected from a design space of $4,320$ combinations using a Markforged Mark Two 3D printer. The processed data formed the input-output set for our proposed model. We compared the results with those from commonly used machine learning models, including feedforward neural networks, Kolmogorov-Arnold networks, XGBoost, CatBoost, and random forests.
Our model achieved the lowest overall test error (MAPE = $12.33\%$) and showed statistically significant improvements over the baseline wide and deep neural network for several target variables (paired t-tests, $p \leq 0.05$). SHapley Additive exPlanations (SHAP) analysis revealed that reinforcement strategy was the major influence on mechanical performance. Overall, this study demonstrates that the integration of LHS and SE-WDNN enables interpretable and sample-efficient multi-target predictions, guiding parameter selection in CFRC-AM with a balance between mechanical behavior and manufacturing metrics.
\end{abstract}

\end{frontmatter}

\section{Introduction}
The increasing need for lightweight, high-performance components in industries such as biomedical, automotive, and aerospace engineering has prompted strong interest in additive manufacturing (AM), which offers unprecedented design freedom by allowing layer-by-layer fabrication of complex geometries from digital designs \cite{Dutra2019, parvaresh}.
In particular, combining AM with continuous fiber-reinforced composite (CFRC) materials enables the fabrication of lightweight structures with outstanding specific strength and stiffness \cite{PradaParra2024, Yi2025}. 
CFRCs are widely used in both civil and military industries due to their phenomenal mechanical properties relative to traditional polymers. They are steadily replacing metal components in the military sector due to their low weight and superior mechanical properties \cite{Zhang2024-deeplearning, Yi2025}. 
Markforged Inc. has developed a series of advanced, specialized three-dimensional (3D) printers with dual extrusion systems that are capable of embedding continuous fibers in a polymer matrix \cite{Lupone2022}. 

Meanwhile, studies have shown that the performance of CFRC-AM is strongly influenced by manufacturing parameters, including fiber type and orientation, number of reinforced layers, concentric rings, infill pattern, fill density, and matrix selection (e.g., nylon vs. onyx) \cite{Zhang2020, Yi2025, RafficNoorMohamed2025, Fetecau2025, Naranjo-Lozada2019, ElEssawi2024}. 
This variability with material heterogeneity and environmental effects \cite{Guo2025, Tian2025, mosadegh}, results in a combinatorial design space where exhaustive testing is impractical and finite element models often fail to capture the coupling among microstructure, process conditions, and performance \cite{JUSTO2018537, Leon-Becerra}.
Reliance on traditional rules such as the rule of mixtures (ROM) \cite{CHABAUD201994, van20153d} in CFRC-AM frequently results in an underestimate by misrepresenting thermoplastic-matrix behavior, especially in the transverse direction \cite{Lupone2022, Dutra2019}. These limitations motivate a shift toward advanced modeling.

Researchers have utilized several machine learning (ML) methods, including deep learning models, such as convolutional neural networks (CNNs), long short-term memory networks (LSTMs), transformer networks, and convolutional autoencoders (CAEs), to make machining quality predictions, recognize AM defects \cite{Akhavan2024, Bhandari2021}, and to improve the design and manufacturing of CFRC-AM. In Table \ref{tab:cfrcml-main}, several ML studies on CFRC-AM  prediction are summarized, including dataset origin, size, features, target variables, and models. Most CFRC-AM studies employ supervised learning to predict tensile properties from a limited set of process and fiber-architecture descriptors, typically using small datasets. In one study, Petcharat et al. \cite{Petcharat2023} use two design variables in a Kriging Surrogate model coupled with NSGA-II to predict mechanical properties. Parra et al. \cite{PradaParra2024} compiles $127$ specimens with eight fiber-architecture features but relies on conventional regressors.

\begin{table*}[!h]
\centering
\caption{Machine learning studies focused on CFRC-AM.\; 
A literature dataset on \emph{pultruded fiber-reinforced polymer (PFRP, non-AM)} is included for comparison.\;see notes below.} \label{tab:cfrcml-main}

\begingroup
\renewcommand{\arraystretch}{1.2}
\footnotesize
\resizebox{1\linewidth}{!}{
\begin{tabular}{|
  >{\centering\arraybackslash}p{2.2 cm} 
| >{\centering\arraybackslash}p{1.8cm}  
| >{\centering\arraybackslash}p{0.8cm}   
| >{\centering\arraybackslash}p{1.4cm}   
| >{\centering\arraybackslash}p{5 cm}   
| >{\centering\arraybackslash}p{2.7 cm} 
| >{\centering\arraybackslash}p{2.5 cm}
|}
\hline
\textbf{Author [Ref]} & \textbf{Process} & \textbf{Type} & \textbf{Size (In$\times$Out)} & \centering\textbf{Features} & \textbf{Targets} & \textbf{Models} \\ \hline\hline

Kumar et al. \cite{Kumar2025} & CFRC\mbox{-}AM & Exp & 36 (2$\times$1) & Layer height; Infill density & Flexural strength & ANN (Levenberg--Marquardt) \\ \hline

Liu et al. \cite{Liu2022} & PFRP (non\mbox{-}AM) & L &746 (12$\times$2) & Fiber type (G/C/B); Matrix type (E/P/V); Fiber volume fraction; Plate thickness; pH; Relative humidity; Exposure time; Exposure temperature & Residual tensile strength; Residual tensile modulus & XGBoost \\ \hline

Petcharat et al. \cite{Petcharat2023} & CFRC\mbox{-}AM & Exp & 45 (2$\times$2) & Infill density; concentric fiber rings & tensile strength; Elastic modulus & Kriging surrogate; NSGA\mbox{-}II \\ \hline

Parra et al. \cite{PradaParra2024} &CFRC\mbox{-}AM &L &127 (8$\times$2) &Fiber volume fraction; Matrix fill pattern; Matrix fill density; Fiber layout (concentric/isotropic); Fiber deposition angle; Matrix deposition angle; Fiber type; Fiber's Young's modulus and tensile strength & Elastic modulus; tensile strength & Ridge; Bayesian Ridge; Lasso; KNN; CatBoost; DT; RF; SVR \\ \hline
\end{tabular}}

\par\medskip
\parbox{0.98\textwidth}{\footnotesize
\textbf{Dataset Type:} \textbf{Exp} = experimental dataset collected in the study; \textbf{L} = literature-compiled dataset.\quad
\textbf{Material abbreviations:} G/C/B = Glass/Carbon/Basalt; E/P/V = Epoxy/Polyester/Vinyl ester.\quad
\textbf{Model abbreviations:} ANN = Artificial Neural Network; KNN = k-Nearest Neighbors; RF = Random Forest ;
NSGA-II = Non-dominated Sorting Genetic Algorithm II; DT = Decision Tree.
}
\endgroup
\end{table*}

As seen in Table \ref{tab:am-others-ml}, datasets from other AM processes are often larger and more varied, yet they are typically left out of CFRC-AM benchmarks \cite{Liu2022}. In polymer-extrusion AM, Jayasudha et al.\cite{Jayasudha2022} compared common regressors on small fused deposition modeling (FDM) and binder jetting (BJ) sets and generally found XGBoost ahead. In a similarly compact study, Liu et al.\cite{Liu2024} predicted spline warpage from composition and processing variables, weighing tree ensembles against support vector regression (SVR). With two outputs, Deb et al.\cite{Deb2024} used a Taguchi design on polylactic acid (PLA) and saw extremely randomized trees (ERTR) lead for tensile strength, while RF best modeled roughness. Expanding to four responses, Ozkul et al.\cite{zkl2025} varied layer thickness, infill, and nozzle temperature on acrylonitrile butadiene styrene (ABS); analysis of variance (ANOVA) pointed to infill density as the primary driver of mechanical properties and layer thickness for roughness. To reduce labeling effort, Tahamina et al.\cite{Nasrin2023} used active learning to build a larger fused filament fabrication (FFF) set and jointly predict five tensile-curve outputs with linear, kernel, and Gaussian process regression (GPR) models. Shifting to metal AM, Akbari et al.\cite{Akbari2024} compiled a broad multi-target benchmark (1,600 builds across powder bed fusion (PBF) and directed energy deposition (DED)) linking process, material properties, and categorical factors to seven mechanical and surface outcomes. Finally, Zhang et al.\cite{Zhang2024-deeplearning} moved to image-based PBF, training on 1,200 metallographic micrographs to predict tensile strength and hardness, with saliency maps highlighting molten-pool morphology. Overall, FDM studies tend to rely on $2$-$6$ process variables and tree- or kernel-based models, which limits direct transfer to CFRC-AM where fiber-architecture descriptors matter.

\begin{table*}[!ht]
\centering
\caption{Machine learning studies for other additive-manufacturing processes.\; Process names are abbreviated in the table; see notes below.}
\label{tab:am-others-ml}

\begingroup
\renewcommand{\arraystretch}{1}
\footnotesize
\resizebox{1\linewidth}{!}{
\begin{tabular}{|
  >{\centering\arraybackslash}p{2.2 cm}  
| >{\centering\arraybackslash}p{1.8cm}   
| >{\centering\arraybackslash}p{0.8cm}   
| >{\centering\arraybackslash}p{1.4cm}   
| >{\centering\arraybackslash}p{5 cm}    
| >{\centering\arraybackslash}p{2.7 cm}  
| >{\centering\arraybackslash}p{2.5 cm} 
|}
\hline
\textbf{Author [Ref]} & \textbf{Process} & \textbf{Type} & \textbf{Size (In$\times$Out)} & \textbf{Features} & \textbf{Targets} & \textbf{Models} \\
\hline\hline

Akbari et al. \cite{Akbari2024} & MAM &
L &1600 (13$\times$7) &Beam power; Layer thickness; Density; Specific heat; Thermal conductivity; Coefficient of thermal expansion (CTE); Melting temperature; Material (cat.); Machine type (cat.); MAM process (cat.); MAM sub-process (cat.); Specimen orientation (cat.); Post-processing condition (cat.) &Yield strength; tensile strength; Elastic modulus; Elongation at break; Vickers hardness; Rockwell hardness; Surface roughness &RF; GBR; SVR; NN; XGBoost \\\hline

Liu et al. \cite{Liu2024} &FDM &Exp &23 (6$\times$1) &PLA content; Chain extender (ADR4468); Twin-screw extrusion conditions; Die swell ratio; Elastic modulus; Impact strength &Spline warpage &SVR; RF; GBR; XGBoost \\\hline

Deb et al. \cite{Deb2024} &FDM &Exp &27 (5$\times$2) &Build angle; Infill angle; Layer thickness; Printing speed; Nozzle temperature &Surface roughness; tensile strength &GBR; XGBoost; ABR; RF; ERTR; SVR; KNN \\\hline

Jayasudha et al. \cite{Jayasudha2022} &FDM; BJ &L &27 (3$\times$1) and 27 (4$\times$1) &Extrusion temperature; Layer height; Shell thickness; Layer thickness; Build orientation (XY, XZ, YZ) & tensile strength &LR; RF; AdaBoost; GBR; XGBoost \\\hline

\"Ozk\"ul et al. \cite{zkl2025} &FDM &Exp &27 (3$\times$4) &Layer thickness; Infill density; Nozzle temperature &Hardness; Tensile strength; Flexural strength; Surface roughness & KStar; MLP; RF; RL; etc. \\\hline

Tahamina et al. \cite{Nasrin2023} &FFF &Exp &124 (2$\times$5) &Extruder temperature; Layer height &Stress 1; Strain 1; Stress 2; Strain 2; Young's modulus &LR; Ridge; GPR; KNN \\\hline

Ziadia et al. \cite{Ziadia2023} &FDM &Exp &54 (3$\times$3) &Printing temperature; Layer thickness; Printing speed &Tensile strength; Young's modulus; Strain at break &RF; XGBoost; GBR; DT; MLR; Lasso; Ridge; blending ensemble (MLR meta-learner) \\\hline

Kadauw \cite{Kadauw2025} &SLA &Exp &54 (4$\times$4) &Orientation; Lifting speed; Lifting distance; Exposure time &Tensile strength; Yield strength; Shore D hardness; Surface roughness &FFN (MATLAB) \\\hline

Zhang et al. \cite{Zhang2024-deeplearning} &PBF &Exp &6$\times$1 (MLP); 1200 images (CNN) &Porosity; Molten pool area/perimeter; Laser power; Scanning speed; Energy density; 224$\times$224 metallographic images (CNN) &Tensile strength; Vickers hardness &MLP; AlexNet; MPR\mbox{-}Net \\\hline
\end{tabular}}

\par\medskip
\parbox{0.98\textwidth}{\footnotesize
\textbf{Dataset Type:} \textbf{Exp} = experimental dataset collected in the study; \textbf{L} = literature-compiled dataset.\quad
\textbf{Process abbreviations} \textbf{SLA} = Stereolithography.\quad
\textbf{Table shorthand:} cat. = categorical.\quad
\textbf{Model abbreviations:}
GBR = Gradient Boosting Regressor ; LR = Linear Regression; MLR = Multiple Linear Regression; FFN = Feedforward Neural Network; NN = Neural Network; MLP = Multilayer Perceptron.
}

\endgroup
\end{table*}

The literature survey shows that two limitations remain in spite of encouraging results. First, as shown in Table \ref{tab:cfrcml-main}, the majority of CFRC-AM studies only examine small parameter subsets with sample sizes ranging from tens to low hundreds. Second, despite the growing popularity of ML, models are frequently shallow or single-target, and capturing inter-target correlations through multi-target learning is rare in these studies. In addition to the high-dimensional design space, these gaps encourage a more data-efficient and expressive method.

We address the research gap identified in prior CFRC-AM studies by proposing a wide and deep neural network (WDNN) enhanced with a squeeze-and-excitation (SE) module for multi-target prediction of both mechanical and manufacturing metrics. While WDNN and SE modules are established neural network components, their integration has not been applied to CFRC-AM, where the input structure involves a mix of continuous geometric parameters, discrete reinforcement settings, and categorical material choices. The WDNN architecture integrates a shallow “wide” component, which excels at memorizing co-occurrence patterns, with a deep component, which generalizes from feature interactions. This dual structure enables the model to capture both low-order and high-order relationships \cite{wide&deep}, making it well-suited for property prediction in additive manufacturing of complex composites. Recent studies further support the utility of WDNNs in materials modeling: Tang et al.\cite{Tang2022} employed a WDNN with finite-element simulations and sensitivity analysis to quantify the influence of multiscale uncertainties on the tensile behavior of 3D woven composites, while Assaf et al.\cite{Assaf2024} demonstrated that an optimized WDNN outperformed conventional learners in predicting thermal and physical properties from molecular-dynamics descriptors.

Building on this foundation, we further incorporate the SE mechanism to strengthen the model’s ability to emphasize the most informative features. SE networks, introduced originally as lightweight channel-attention modules, recalibrate feature maps adaptively and have shown increasing value in manufacturing and mechanics \cite{SE-mainRef}. For example, Kong et al.\cite{Konglingbao-SE} integrated SE blocks into a YOLOv3 framework with MobileNetV3, yielding significant improvements in precision and recall for in-situ defect detection of printed components. Sun et al.\cite{sunYixun-SEBlock} embedded SE modules into residual convolutional structures to predict local stress distributions in fiber-reinforced composites, achieving more accurate microstructural characterization than standard CNNs. Likewise, Zhang et al. \cite{zhangshentong-SEBlock} demonstrated that SE-enhanced CNNs accelerate and improve property prediction for heterogeneous materials in multiscale mechanical modeling. Collectively, these studies underscore the relevance of SE-enhanced architectures in manufacturing research and motivate our integration of SE modules into WDNNs for CFRC-AM, where capturing subtle yet critical feature-to-feature interactions is essential for reliable defect detection and property prediction.

Our key contributions are summarized as follows:
\begin{enumerate}
\item We propose an innovative SE-WDNN architecture that captures both low-order and high-order feature interactions. It adaptively rescales feature maps. This architecture goes beyond traditional single-target models by predicting multiple mechanical responses alongside production and cost metrics for CFRC-AM, all while using the same feature set.
\item We generate $155$ representative specimens using Latin hypercube sampling (LHS) in a high-dimensional space defined by seven process and material parameters, providing a diverse and systematic exploration of CFRC-AM variability.
\item We integrate SHapley Additive exPlanations (SHAP) to quantify feature importance and provide mechanistic insights into structure-property relationships, thereby identifying the parameters most critical to manufacturing outcomes.
\end{enumerate}

To the best of our knowledge, this work represents the first systematic integration of LHS-driven experimentation with an SE-WDNN framework for multi-target prediction in CFRC-AM processes. The remainder of this paper is organized as follows: Section \ref{Sec:methods} details the materials, manufacturing setup, and experimental design; Section \ref{Sec:WDNN} introduces the SE-WDNN architecture and baseline models; Sections \ref{Sec:results} and \ref{sec:ablation} present the results and feature importance analysis respectively; and Section \ref{Sec:conclusion} provides concluding remarks and future research directions.

\section{Materials and Methods}\label{Sec:methods}
In this section, we explain the manufacturing materials and parameters used, the manufacturing process, the LHS, and the ML model for prediction. 

\subsection{Materials and Manufacturing Parameters of CFRC-AM} 

Several manufacturing parameters must be defined and adjusted before printing the specimens, as they directly affect the quality and mechanical properties of composite parts made using AM. Since it was not feasible to explore the full range of manufacturing parameters in this study, certain parameters (factors)  -  such as {\it top and bottom layers} and {\it wall layers}-  were set as constants for all samples. 

Top and bottom layers refer to the solid layers at the top and bottom surfaces of the printed part. They create the outer boundaries in the z-direction and mainly influence the surface finish and structural integrity. Wall layers determine the number of vertical perimeter walls, or shells, printed around each layer. A higher wall count increases edge strength and dimensional accuracy, but it also uses more material and takes more time to build. In this study, we kept these two parameters constant for all specimens. On the Markforged Mark Two, other process conditions, like nozzle and bed temperature, print speed, extrusion flow rate, and fiber feed are controlled by the printer's firmware. These conditions do not vary between builds and were not included as input features. Instead, we systematically varied the seven manufacturing parameters listed in Table~\ref{tab:manwufparam} to define the design space we explored in this work.

\begin{table}[ht]
\centering
\caption{Manufacturing parameters and levels used to fabricate the CFRC-AM specimens. The parameters were explored at the listed levels, while constants were held fixed across all builds. }\label{tab:manwufparam}
\begin{adjustbox}{max width=\columnwidth}
\begin{tabular}{|l| c|}
\hline
\textbf{Manufacturing Parameter} & \textbf{Levels} \\
\hline \hline
Plastic Types                  & Onyx, Nylon \\
\hline
Fiber Types                    & Carbon Fiber, Fiberglass, Kevlar \\
\hline
Fill Density (\%)             & 28, 36, 44, 52 \\
\hline
Infill Pattern                & Gyroid, Triangular, Hexagonal, Rectangular \\
\hline
Number of Reinforced Layers   & 2, 4, 6, 8, 10 \\
\hline
Fiber Orientation ($^\circ$)         & 0, 45, 90 \\
\hline
Concentric Rings              & 1, 2, 3 \\
\hline
 Top and bottom layers& $5$\\
\hline
Wall layers & $2$\\
 \hline
\end{tabular}
\end{adjustbox}
\end{table}

\begin{figure}[!ht]
\centering

\begin{subfigure}[t]{0.494\linewidth}
    \centering
    \includegraphics[height=6cm, width=\linewidth]{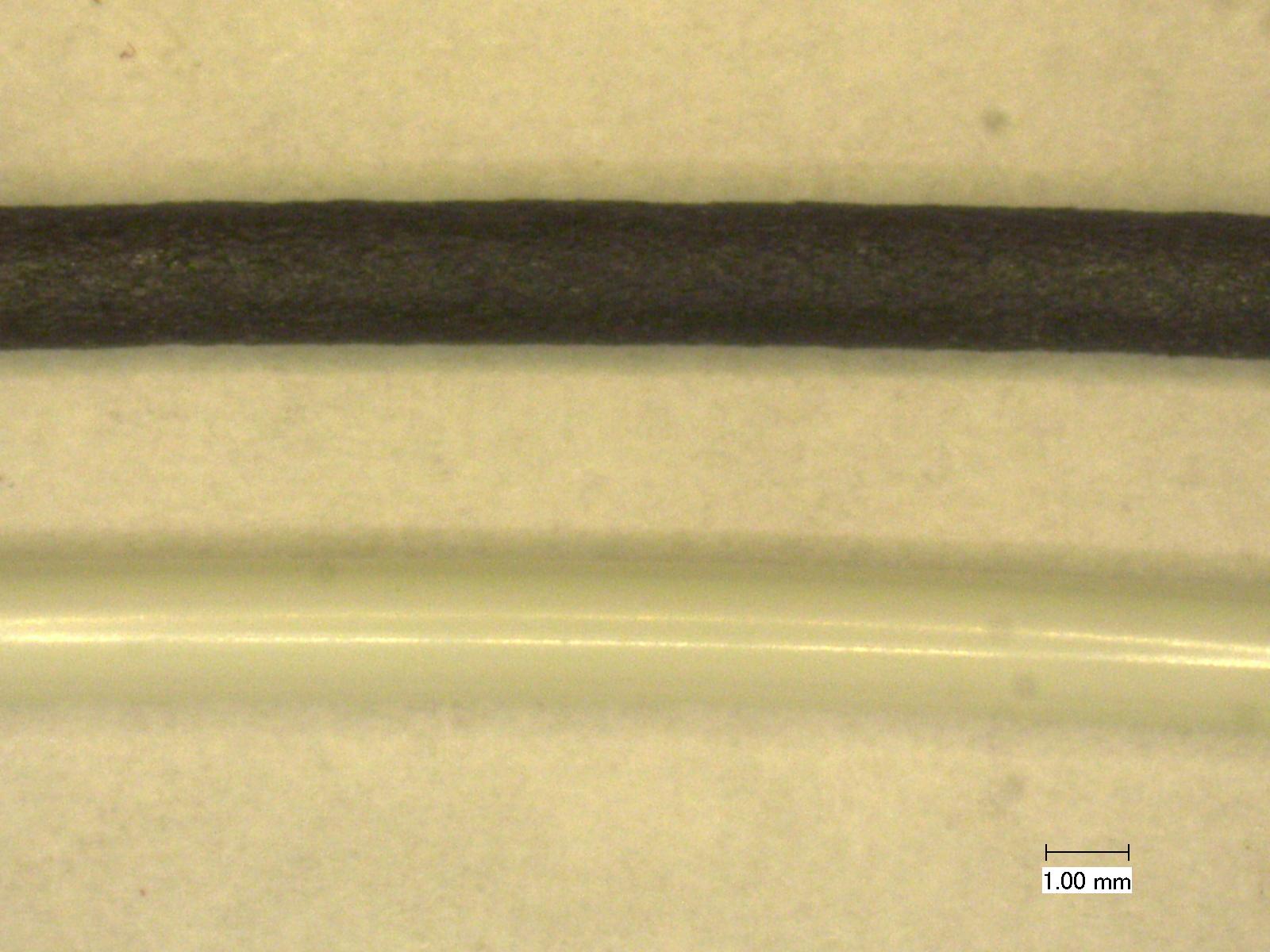}
    \caption{Microscopic image of the thermoplastic matrix materials used in this work. From top to bottom: Onyx and Nylon filaments}
    \label{fig:plastic_type}
\end{subfigure}
\hfill
\begin{subfigure}[t]{0.494\linewidth}
    \centering
    \includegraphics[height=6cm, width=\linewidth]{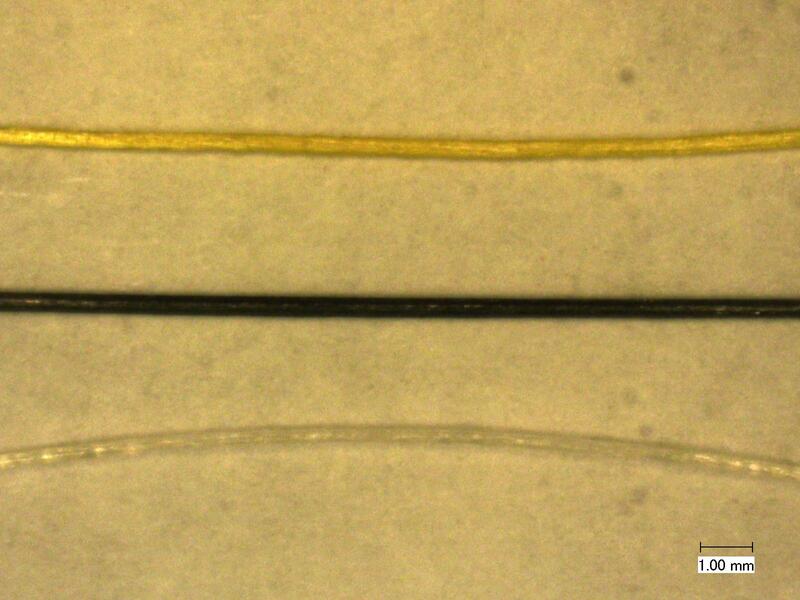}
    \caption{Microscopic images of the continuous-fiber reinforcements. From top to bottom: Kevlar (yellow), carbon fiber (black), and fiberglass (transparent). }
    \label{fig:fiber_type}
\end{subfigure}
\caption{(a) Microscopic images of the thermoplastic matrix filaments used in this work (b) Microscopic images of the continuous-fiber reinforcements}
\end{figure}

\paragraph{Plastic and Fiber Types }
A composite material is made of two types, which are the matrix (plastic) and the reinforcement (fiber). {\it Plastic} type is the thermoplastic polymer used as the matrix material in the composite material-specifically Nylon and Onyx. The matrix choice has a noticeable impact on the composite's ductility, heat resistance, printability, and overall mechanical performance. Nylon offers flexibility, solid wear resistance, and strong fatigue properties. Onyx, which is a microcarbon-filled nylon, is known for its relatively high toughness and flexural strength, making it a strong candidate for components where strength and precision are critical. Figure \ref{fig:plastic_type} shows microscopic images of both plastic types used in this work. 

The {\it reinforcement} phase is critical to the overall performance of composite materials. Carbon, Fiberglass, and Kevlar are three types of continuous fibers considered in this study, each with distinct strengths for engineering applications. Carbon fiber is commonly used in applications requiring very high tensile strength, stiffness, strength-to-weight ratio, and low weight, making it ideal for aerospace and automotive components. Kevlar is less stiff but is known for its excellent impact energy absorption and ultra-toughness, making it suitable where impact resistance and energy absorption are essential. Fiberglass, though less strong and stiff than carbon fiber, is more cost-effective and provides good impact resistance, making it practical for large-scale or budget-sensitive projects. The microscopic images in Figure \ref{fig:fiber_type} illustrate the different surface textures and structures of these three continuous fibers.

As noted by Markforged Inc., AM of composites using Onyx as the matrix and Kevlar as the reinforcement produces a material resistant to catastrophic fracture, making it suitable for demanding applications or parts subjected to repeated loading. The physical and mechanical properties of these matrix and reinforcement materials, as reported in the Markforged Inc. material datasheet \cite{MarkforgedInc2025}, are provided in Appendix \ref{AppA}.

\begin{figure}[!ht]
\centering

\begin{subfigure}[t]{0.494\linewidth}
    \centering
    \includegraphics[width=\linewidth]{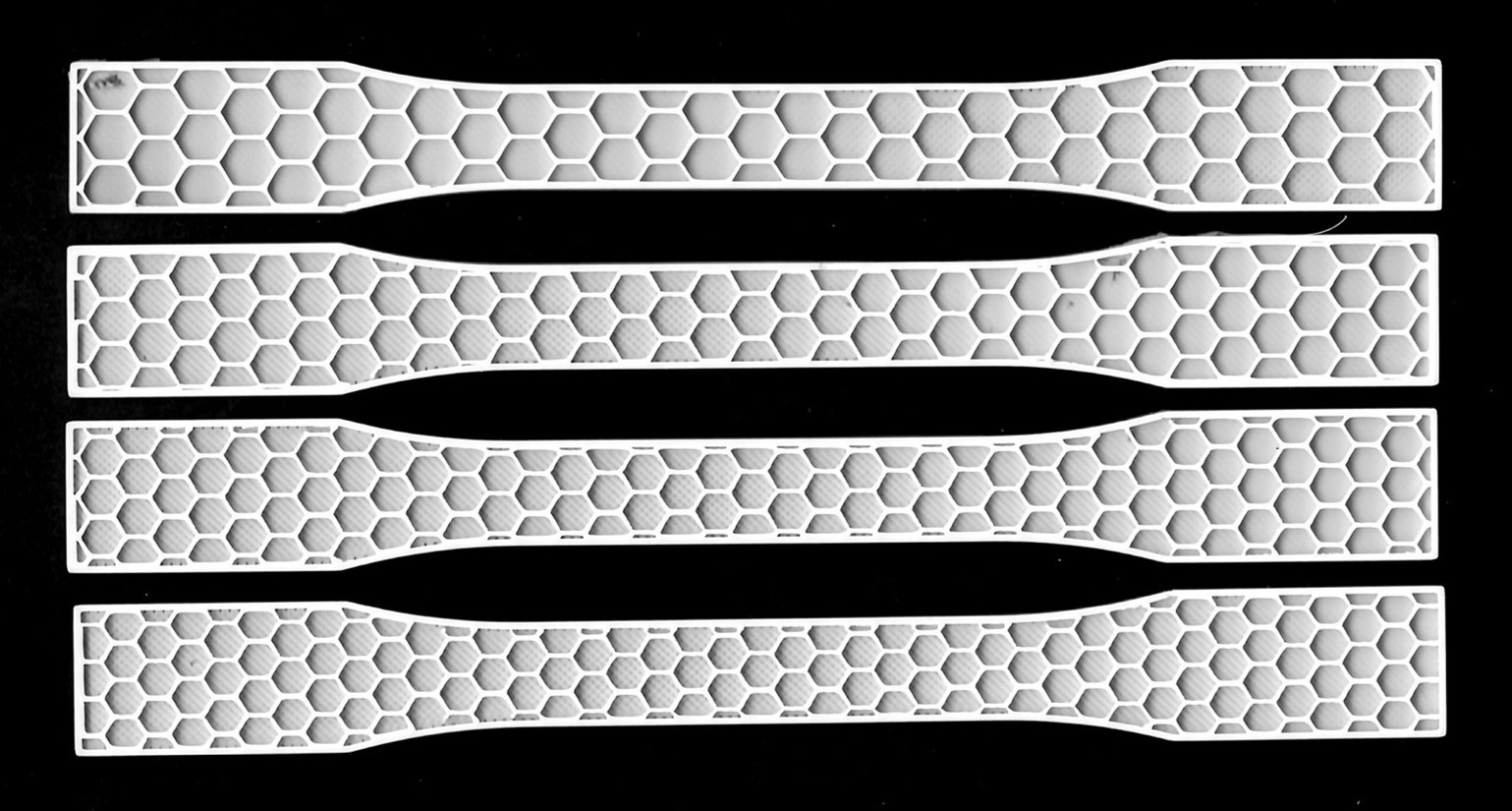}
    \caption{Fill-density levels for CFRC-AM specimens with hexagonal infill; from top to bottom: 28\%, 36\%, 44\%, and 52\%. }
    \label{fig:filldensity}
\end{subfigure}
\hfill
\begin{subfigure}[t]{0.494\linewidth}
        \includegraphics[width=\linewidth]{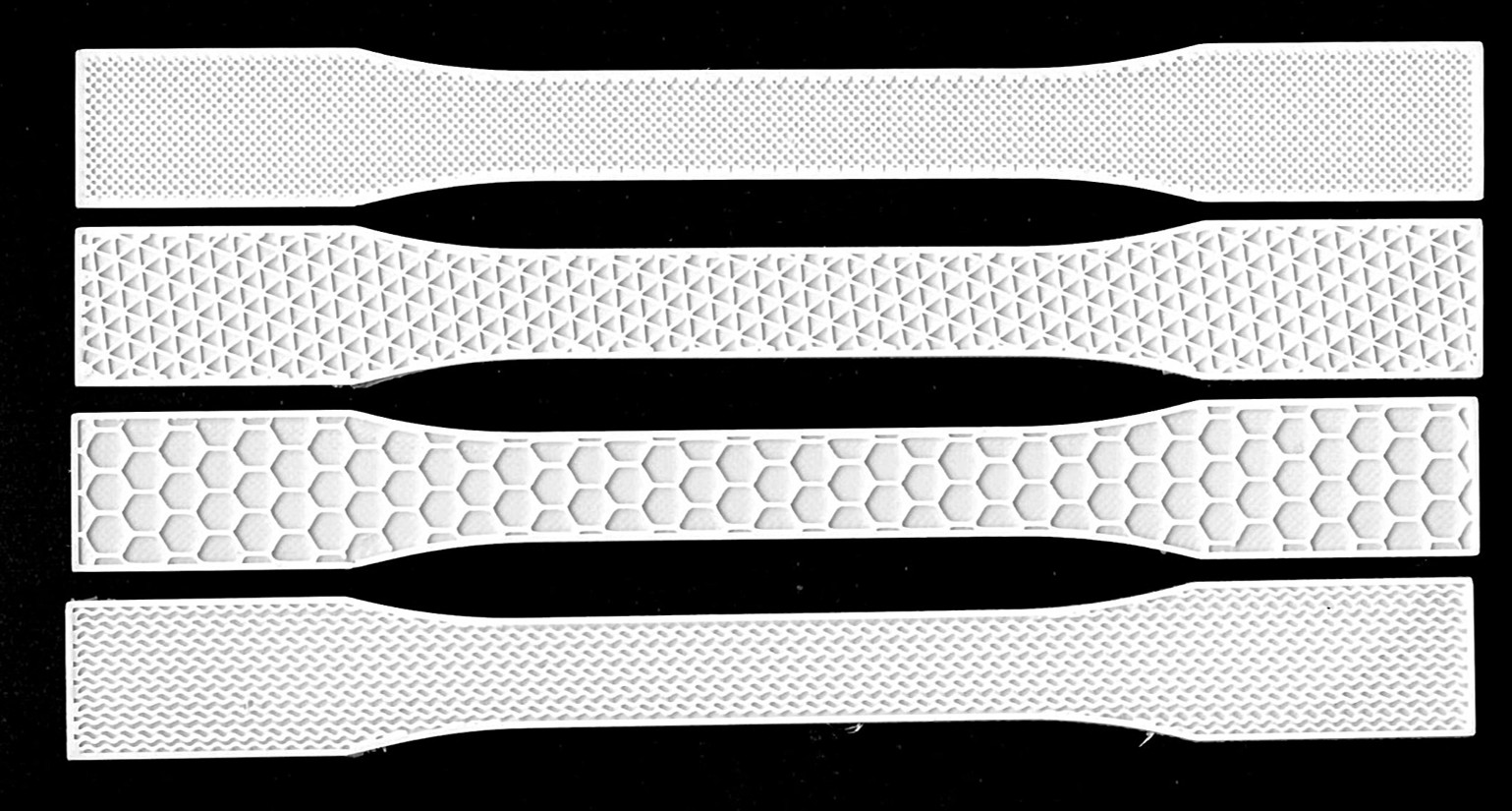}
    \caption{Representative infill patterns for CFRC-AM; from top to bottom: rectangular, triangular, hexagonal, and gyroid. }
    \label{fig:infillPattern}
\end{subfigure}
\caption{(a) Fill density levels for CFRC-AM specimens with hexagonal infill; from 28\% to 52\% (b) Representative infill patterns for CFRC-AM. }
\end{figure}

\paragraph{Fill Density $(\%)$}
In CFRC-AM, fill density refers to the percentage of internal volume filled with polymer matrix material. Higher fill density improves load transfer between fibers and the matrix, thereby enhancing stiffness, strength, and overall mechanical stability. However, it also increases weight, manufacturing time, material consumption, and cost. Conversely, lower fill density reduces component mass and fabrication costs but may compromise composite strength, structural integrity, and fatigue resistance. Thus, selecting an appropriate fill density is essential to balance performance and manufacturing efficiency in CFRC-AM. Illustrative examples of the fill-density are shown in Figure \ref{fig:filldensity}.

\paragraph{Infill Pattern}
The infill pattern specifies the geometric configuration used to fill the internal volume of any printed part, with common patterns being gyroid, triangular, and honeycomb. 
This parameter is important because it influences load distribution and stress transfer between layers, thereby affecting stiffness, deformation, and anisotropy. It also impacts build time and material consumption. 
Thus, the choice of infill pattern plays a dual role in determining both the mechanical performance and manufacturing efficiency of CFRC-AM. Representative infill options used in this study are shown in Figure \ref{fig:infillPattern}.

\paragraph{Number of Reinforced layers}
The number of layers reinforced with continuous fibers significantly influences the mechanical performance of composite parts. 
Increasing reinforcement raises the fiber volume fraction, leading to improved mechanical properties, better stress distribution, and delayed crack formation. 
However, beyond a certain point, the benefits plateau while weight and production costs continue to rise. Therefore, careful optimization of the amount of reinforced layers is essential for achieving balanced performance in fiber-reinforced composites \cite{Dong2024}.

\paragraph{Fiber Orientation ($^\circ$)}

\begin{figure}[!ht]
\centering

\begin{subfigure}[t]{0.494\linewidth}
    \centering
    \includegraphics[width=\linewidth]{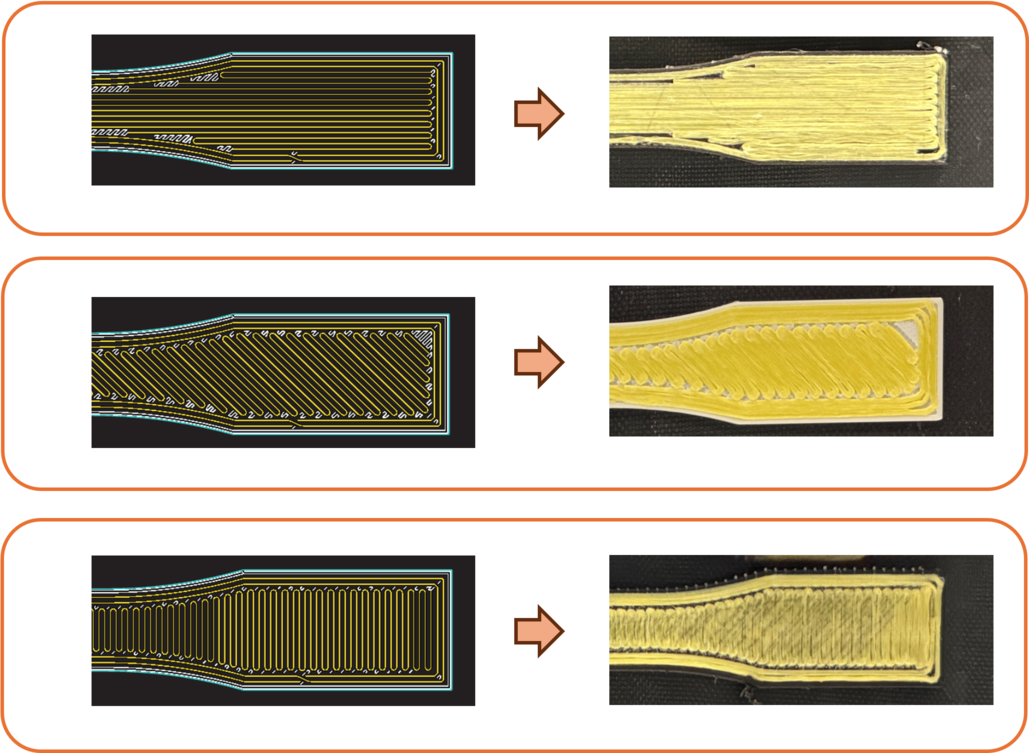}
    \caption{Fiber orientation configurations in CFRC-AMs. Schematic tool paths (left) and corresponding printed specimens (right). The in-plane fiber angle $\theta$ is measured relative to the specimen's longitudinal (loading) axis; from top to bottom: $0^\circ$, $45^\circ$, and $90^\circ$. }
    \label{fig:fiber orientation}
\end{subfigure}
\hfill
\begin{subfigure}[t]{0.494\linewidth}
     \centering
    \includegraphics[width=\linewidth]{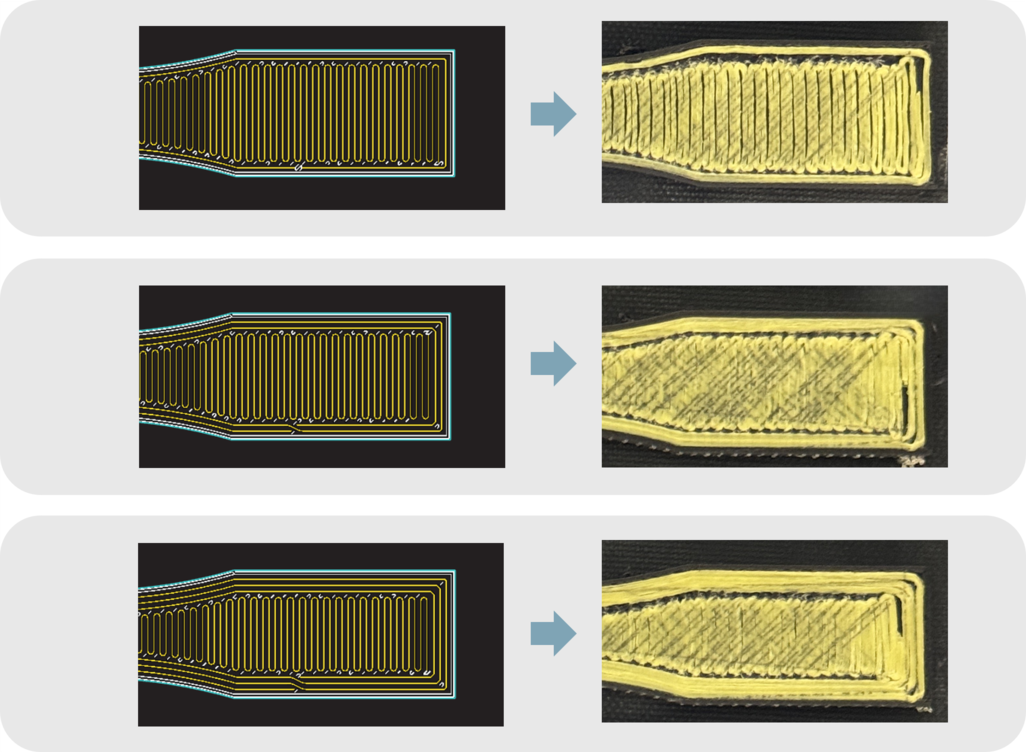}
    \caption{Concentric ring configurations in CFRC-AMs. Schematic tool paths (left) and corresponding printed specimens (right), all with $90^\circ$ fiber orientation. From top to bottom, the ring count increases from one to three concentric rings. }
    \label{fig:concentricRings}
\end{subfigure}
\caption{(a) Fiber orientation configurations, and (b) concentric ring configurations in CFRC-AM specimens. Each panel shows schematic tool paths on the left and the printed specimens on the right. }
\end{figure}

Fiber orientation refers to the angle at which continuous fibers are placed in each printed layer (Figure \ref{fig:fiber orientation}). It strongly influences the mechanical response of CFRC-AM components. 
When fibers are aligned with the applied load, stiffness and tensile strength are maximized, whereas off-axis orientations lead to shear- and matrix-dominated failure modes. 
Thus, proper selection of fiber angle is critical for tailoring strength, stiffness, and failure behavior \cite{parmiggiani2021effect}.

\paragraph{Concentric Rings}
Concentric rings refer to the number of closed continuous-fiber loops placed along the perimeter of each printed layer (Figure \ref{fig:concentricRings}). Increasing the ring count enhances edge stiffness and strength by adding reinforcement and improving load transfer, but also increases fiber consumption and build time.

\subsection{Manufacturing Process}
\begin{figure}[!ht]
    \centering
    \includegraphics[width=\linewidth, frame]{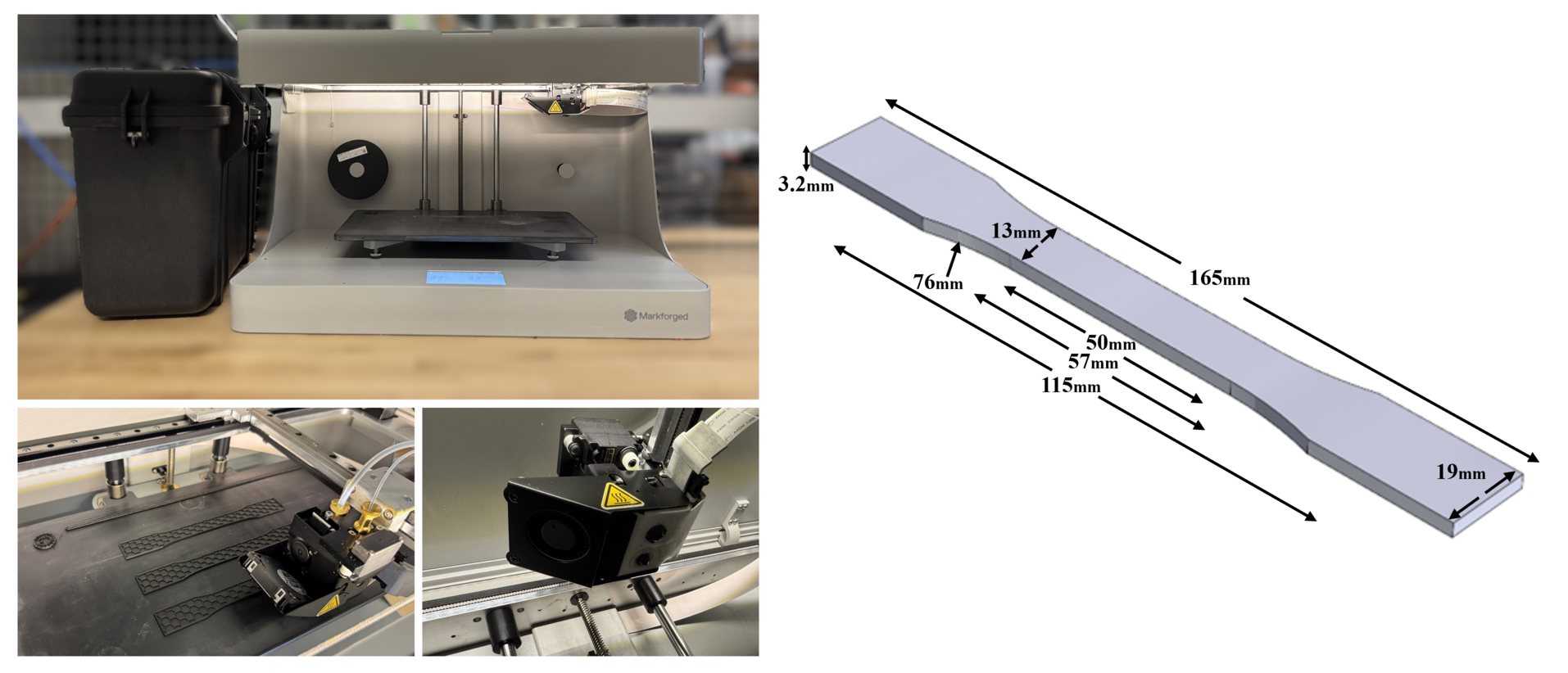}
    \caption{Markforged Mark Two composite 3D printing setup and test-specimen geometry. Left: the printer used in this study, which combines fused filament fabrication with continuous-fiber reinforcement via a dual-nozzle printhead (insets show the build plate and tool head). Right: CAD model of the ASTM D638 Type I tensile specimen with key dimensions indicated. }
    \label{fig:processanddimensions}
\end{figure}

The specimens were manufactured using a Markforged Mark Two composite 3D printer, which combines fused filament fabrication (FFF) with continuous fiber reinforcement (CFR) technology. As shown in Figure \ref{fig:processanddimensions}, the print head has two nozzles that enable the deposition of continuous fiber strands into a thermoplastic matrix layer by layer. 

As seen in Figure \ref{fig:processanddimensions}, specimens were designed according to ASTM D638-14, Type I geometry \cite{ASTMD638}. ASTM D638 states that reinforced plastics and highly orthotropic laminates can be tested using this standard if the specimen meets the Type I dimensions. This method is commonly used in the CFRC-AM literature. In particular, Naranjo-Lozada et al.\cite{Naranjo-Lozada2019}, El Essawi et al. \cite{ElEssawi2024}, and Mohammadizadeh et al. \cite{Mohammadizadeh2021} made CFRC-AM specimens following ASTM D638 Type I and reported tensile modulus and strength.
In CFRC-AM, tensile coupons are made individually in their final shape instead of being cut from larger, flat laminates. Printing the ASTM D638 Type I specimen directly keeps the as-printed fiber paths, infill patterns, and void structure intact without needing extra machining or tab bonding. On the other hand, ASTM D3039 \cite{ASTMD3039} is the best test method for continuous-fiber laminates. It needs straight-sided rectangular coupons with bonded end tabs to prevent grip failures. Although ASTM D3039 offers a more accurate description of composite behavior, its preparation steps are not well-suited to the discrete, near-net-shape geometries created by CFRC-AM process using Markforged 3D printer. For this reason, in line with previous CFRC-AM studies and the limits of the printing process, ASTM D638 Type I was chosen as the right tensile test standard for the configuration investigated in this work.

\subsection{Design of Experiments}
The specified factors and levels yield $4,320$ unique parameter combinations. A full factorial is infeasible under time and resource constraints, and simple random sampling risks leaving large regions of the design space unexplored. In CFRC-AM, the design spans multiple categorical and numerical factors (e.g. , plastic and fiber type, infill pattern, orientation, fill density), so we propose LHS to obtain stratified, space-filling coverage with a limited number of trials. LHS guarantees stratification along each input dimension, reducing variance relative to simple random sampling and providing balanced coverage that better captures factor effects and interactions. 

\paragraph{LHS formulation.}  
Let the design space be defined by $d$ input parameters $X_1,\dots,X_d$, each normalized to $[0,1]$. The goal is to generate $N$ design points $\{{z}^{(1)},\dots,{z}^{(N)}\}\in[0,1]^d$ such that each marginal distribution is uniformly stratified. For each dimension $i$, the interval $[0,1]$ is partitioned into $N$ intervals
\begin{equation}
I_{i,k}=\bigg[\tfrac{k-1}{N},\tfrac{k}{N}\bigg), \quad k=1,\dots,N. 
\end{equation}
Draw one sample $z_{i,k}\in I_{i,k}$ per interval, then permute the $N$ intervals independently for each dimension using a random permutation $\pi_i$. The $j$-th design point is then
\begin{equation}
{z}^{(j)} = \big(z_{1,\pi_1(j)},\,z_{2,\pi_2(j)},\,\dots,\,z_{d,\pi_d(j)}\big), \quad j=1,\dots,N. 
\end{equation}
By construction, each interval appears exactly once in every dimension, ensuring stratified one-dimensional projections \cite{Wei-Liem}. In our design, about $60$ samples suffice to cover each factor level at least once, while $>120$ are needed to capture most pairwise interactions. To account for higher-order effects and ensure adequate density, we selected $N=155$ LHS points.

\paragraph{Implementation details (mixed factors, optimization, diagnostics). }  
We first generate the normalized design ${Z}=[{z}^{(1)},\dots,{z}^{(N)}]^\top\in[0,1]^{N\times d}$, then map to physical factor levels: for continuous/ordinal parameters, $x_i^{(j)} = F_i^{-1}(z_i^{(j)})$, where $F_i$ is the specified or empirical CDF. For discrete ordered factors, $x_i^{(j)}=\ell_{i,k}$ with $k=1+\min(m_i-1,\lfloor m_i z_i^{(j)}\rfloor)$. For  
categorical unordered factors, assign levels so each appears $\approx N/m_i$ times (balanced binning). 

To enhance space-filling, we optimize ${Z}$ via pairwise swaps to maximize
\begin{align}
S({Z})=\alpha\,\phi_{\min}({Z})-\beta\,\Phi_q({Z})-\gamma\,\rho_{\mathrm{sum}}({Z}),
\end{align}
where $\phi_{\min}$ is the minimum inter-point distance, $\Phi_q$ the Morris -Mitchell criterion, and $\rho_{\mathrm{sum}}$ the sum of absolute column correlations. From $R$ replicates, the design with maximal $S$ is retained. Manufacturing constraints $\mathcal{C}$ are handled by either repair (nearest feasible level) or resampling.

\subsection{Mechanical Testing}

Tensile testing was conducted at room temperature on an \texttt{Instron 5967} with a $30\,kN$ load cell, following the ASTM D638 standard \cite{ASTMD638}. A constant crosshead speed of $5\,mm/min$ was used. 
From the resulting engineering stress and strain curves, we determined the main mechanical parameters: tensile strength, yield strength, Young's modulus, strain at break, and toughness. The slope between two points on the stress-strain curve at $\sigma_1$ at $\varepsilon_1=0.05\%$ and  $\sigma_2$ at $\varepsilon_1=0.25\%$ is the Young modulus. Also, by drawing a line parallel to the Young's modulus line, but offset by a strain of $ 0.2\%$, and finding the intersection of this line and the stress-strain curve, determines the yield strength \cite{Naranjo-Lozada2019}. The maximum stress value in the stress-strain curve is tensile strength. Also, the strain at break is a strain where the sample is pulled until it suddenly fractures. 
In polymer composites, the fracture may occur in multiple stages with stress decreasing gradually; in these cases, strain at break was taken at the last recorded strain before the stress decreased to $10\%$ or less of the tensile strength \cite{standard2019plastics}. The area under the engineering stress-strain curve is Toughness, which is calculated by trapezoidal rule using the equation \ref{eq:toughness} \cite{Yilmaz2019}: 
\begin{align}
  \text{Toughness} \approx \sum_{i=1}^{n-1} \frac{1}{2} \left( \sigma_i + \sigma_{i+1} \right) \left( \varepsilon_{i+1} - \varepsilon_i \right)
  \label{eq:toughness}
\end{align}

Also, the resilience is the amount of energy that material can absorb and release upon unloading without having permanent deformation \cite{Lin}, and is calculated using the equation \ref{eq:resilience} \cite{ElEssawi2024}: 
\begin{align}
  \text{Resilience} = \frac{1}{2} \cdot \frac{(\text{Yield Strength})^2}{\text{Young's Modulus}}
  \label{eq:resilience}
\end{align}

To assess experimental repeatability within practical limits, we selected 35 out of the 155 LHS-generated parameter combinations for replication. Each of these 35 conditions was printed and tested twice, leading to a total of 70 tensile tests. Fully replicating all 155 conditions was not possible because of the high material use, long build times, and costs linked to CFRC-AM process. For the 35 replicated conditions, we averaged the mechanical properties to create the final dataset for model training.

In CFRC-AM, there are unavoidable sources of process variability, including local voids and small differences in layer bonding. The Markforged Mark Two reduces this variability to some extent with its closed, pre-calibrated control of fiber routing and thermal conditions, but it cannot eliminate it completely \cite{PARKER2023111505}. The replicated tests and CV analysis provide an empirical estimate of this remaining process noise and make sure that the dataset represents realistic manufacturing uncertainty.

After printing the material and parameter combinations defined by LHS, we analyzed each of the eight targets, with their summary statistics presented in Table \ref{tab:target_summary}. 

\begin{table}[ht]
\centering
\caption{Summary statistics (Mean, Standard deviation (Std Dev), Minimum value (Min), and Maximum value (Max)) of the target parameters used for model evaluation, which include the mechanical and manufacturing properties of CFRC. The statistics are calculated on the final modeling dataset after averaging the two measurements for each of the 35 replicated processing conditions. They describe the variability across all 155 process configurations in the design space. }\label{tab:target_summary}
\begin{adjustbox}{max width=\columnwidth}
\begin{tabular}{| l| c| c| c |c|}
\hline
\textbf{Target Parameter} & \textbf{Mean} & \textbf{Std Dev} & \textbf{Min} & \textbf{Max} \\
\hline \hline
Toughness (MPa) & 1.52  & 1.09  & 0.20  & 5.36 \\ \hline
Strain at Break (\%) & 4.17  & 7.52  & 1.23  & 91.21 \\ \hline
Young's Modulus (GPa) & 4.28  & 3.36  & 0.58  & 17.03 \\ \hline
Yield Strength (MPa) & 71.02 & 48.98 & 12.95 & 216.70 \\ \hline
Tensile Strength (MPa) & 88.60 & 59.80 & 19.24 & 274.14 \\ \hline
Build Time (min) & 97.01 & 11.97 & 69.00 & 127.00 \\ \hline
Weight (g) & 8.04  & 0.89  & 5.88  & 10.24 \\ \hline
Material Cost (\$) & 4.35  & 1.93  & 1.76  & 9.30 \\\hline
\end{tabular}
\end{adjustbox}
\end{table}

\section{Wide and deep neural network with Squeeze-and-Excitation Prediction Model}\label{Sec:WDNN}

\begin{figure*}[ht]
    \centering
    \includegraphics[width=\linewidth, frame]{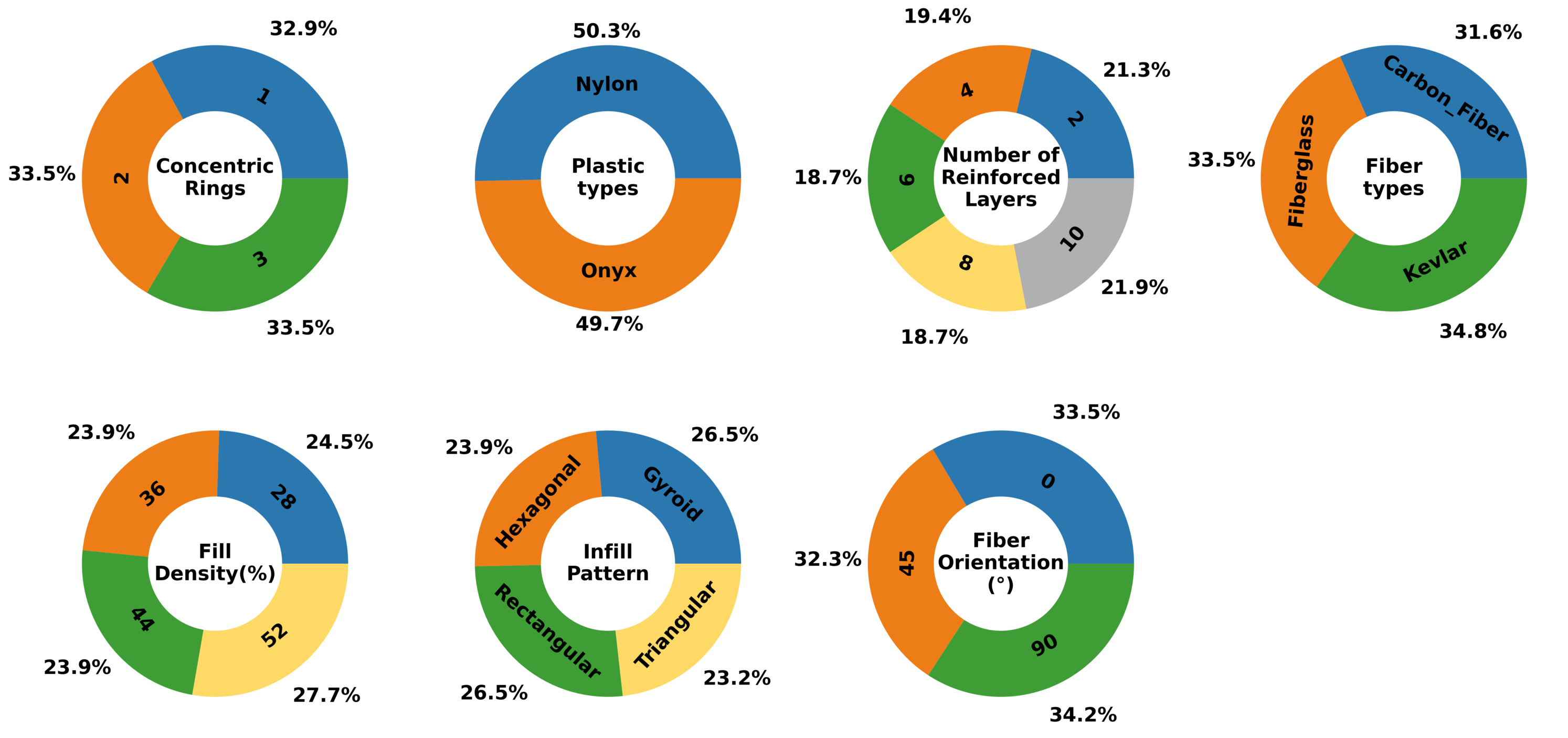}
    \caption{Distribution of the manufacturing parameters used as input features for CFRC-AM specimens, including categorical parameters (plastic type, fiber type, infill pattern) and numerical parameters (fiber orientation, concentric rings, fill density, number of reinforced layers). }
    \label{fig:inputdistr}
\end{figure*}

The processed dataset consists of $155$ specimens characterized by their manufacturing parameters, namely: {\it Plastic types, Fiber Types, Fill  Density, Infill Pattern, Number of Reinforced Layers, Fiber Orientation, and Concentric Rings}. Figure \ref{fig:inputdistr} illustrates the distribution of these parameters and highlights the coverage of different factor levels in the experimental dataset. Given these parameters as input features, our research goal is to predict the economic properties-{\it Weight, Build Time, Material Cost}- and mechanical properties, including {\it Young's modulus, Yield Strength, Tensile strength, Strain at Break, Toughness, and Resilience. } 

\begin{figure*}[ht]
    \centering    \includegraphics[width=\linewidth, frame]{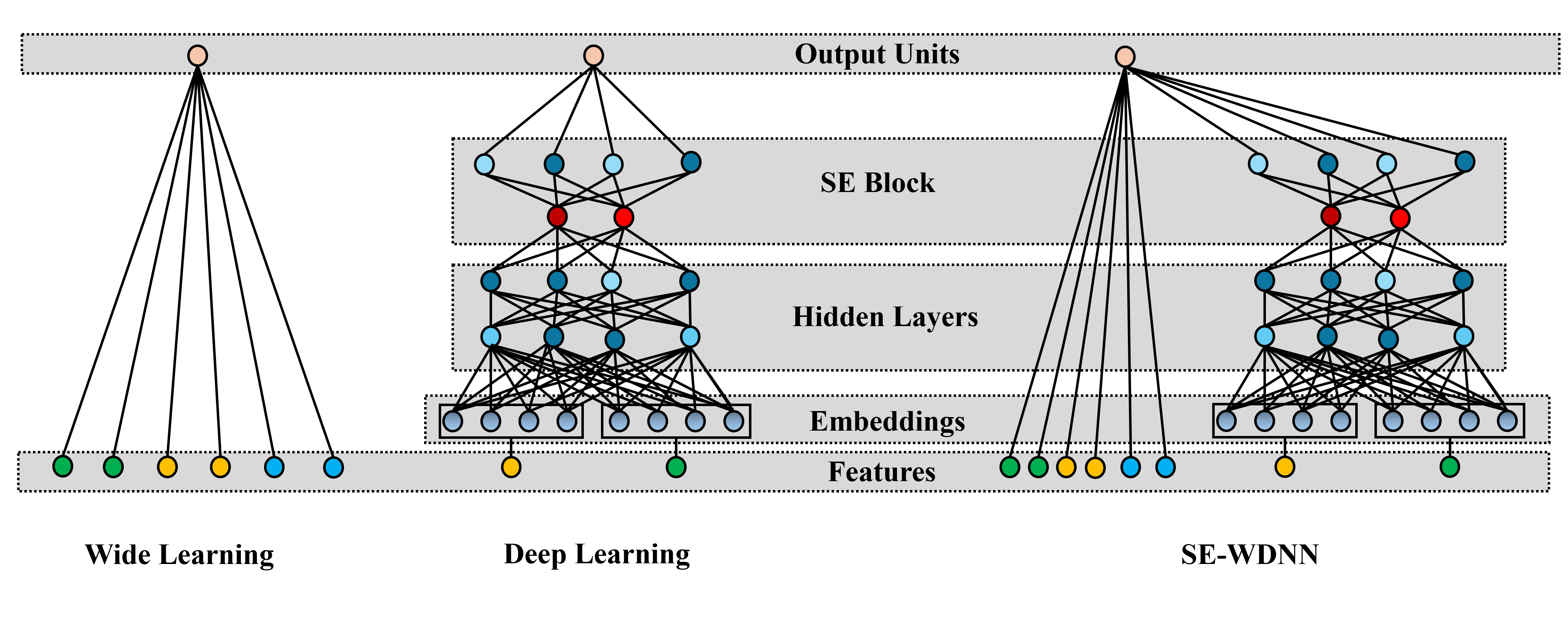}
    \caption{Models architecture: Wide learning; Deep network with a squeeze-and-excitation block; Proposed SE-WDNN model in this study. }
    \label{fig:SE-WDNN}
\end{figure*}

As seen in \ref{fig:SE-WDNN}, we propose a hybrid wide and deep neural network enhanced with a squeeze-and-excitation block to predict multiple mechanical and economic properties of fiber-reinforced polymer samples. The workflow comprises three main stages: feature representation, network architecture design, and model training and evaluation.

\subsection{Feature Representation}
Let the original input-output dataset consisting of $N$ samples be denoted by $\{(x^{(i)},y^{(i)})\}_{i=1}^N$, and each input sample is further partitioned into numerical feature $x_{\mathrm{num}}^{(i)}$ and categorical features $x_{\mathrm{cat}}^{(i)}$, denoted by
\begin{align}
x^{(i)} = \bigl[x_{\mathrm{num}}^{(i)},\,x_{\mathrm{cat}}^{(i)}\bigr]
\quad
x_{\mathrm{num}}^{(i)} \in \mathbb{R}^p,\quad
x_{\mathrm{cat}}^{(i)}\in \{1,\dots,V\}^q. 
\end{align} 

The numerical features are standardized, such that we have 
  \begin{align}
      \tilde{x}_{\mathrm{num},j}^{(i)} = 
      \frac{x_{\mathrm{num},j}^{(i)} - \mu_j}{\sigma_j},
      \quad j=1,\dots,p,
    \end{align}
    where $\mu_j$ and $\sigma_j$ are the sample mean and standard deviation.

We construct two parallel inputs from the transformations:h
\paragraph{Deep branch:}  
    For each categorical feature $x_{\mathrm{cat},j}$, an embedding layer
    $\mathrm{Emb}_j : \{0,\dots,V_j-1\}\to\mathbb{R}^{e_j}$. Denote
    ${e}^{(i)}\in\mathbb{R}^E$, $E = \sum_j e_j$. Concatenate with the
    standardized numerical vector $\tilde{{x}}_{\mathrm{num}}^{(i)}\in\mathbb{R}^{p}$ to form
    \begin{align}
      {z}^{(i)} = \bigl[{e}^{(i)};\,\tilde{{x}}_{\mathrm{num}}^{(i)}\bigr]\in\mathbb{R}^{E + p}. 
    \end{align}
\paragraph{Wide branch:}  
    Sparse one‐hot vector ${w}^{(i)}\in\{0,1\}^d$.

\subsection{SE-WDNN Architecture}
The deep input ${z}$ is transformed by two fully-connected layers:
\begin{align}
  {h}_1 &= \mathrm{Dropout}\bigl(\mathrm{BatchNorm}(\mathrm{ReLU}(W_1\,{z} + {b}_1))\bigr)
    \in\mathbb{R}^{128},\\
  {h}_2 &= \mathrm{Dropout}\bigl(\mathrm{ReLU}(W_2\,{h}_1 + {b}_2)\bigr)
    \in\mathbb{R}^{64}. 
\end{align}

To enable feature-wise adaptive re-weighting, we introduce an SE block after the fully connected layers:
\begin{align}
  {s} &= \mathrm{ReLU}\bigl(W_s\,{h}_2 + {b}_s\bigr)\in\mathbb{R}^{r},\quad r = \lfloor 64/4\rfloor = 16,\\
  {g} &= \sigma\bigl(W_e\,{s} + {b}_e\bigr)\in (0,1)^{64},\\
  {d} &= {h}_2 \odot {g}\in\mathbb{R}^{64}. 
\end{align}
This produces the final deep representation ${d}^{(i)}$. The wide branch applies a dense layer:
\begin{align}
  z_{\mathrm{wide}}^{(i)} = W_w\,{w}^{(i)} + {b}_w \;\in\mathbb{R}^K,
\end{align}
where $K$ is the number of outputs. 
We then concatenate ${d}^{(i)}$ and $z_{\mathrm{wide}}^{(i)}$ and pass through a final dense layer of the output shape:
\begin{align}
  \hat{{y}}^{(i)} = W_o\,[\,{d}^{(i)};\,z_{\mathrm{wide}}^{(i)}\,] + {b}_o
  \;\in\mathbb{R}^K. 
\end{align}

\subsection{Training and Evaluation}
In this study, we considered eight target parameters, which were measured on a different scale (for example, build time in minutes and weight in grams). Hence, an evaluation metric independent of scale was needed to allow fair comparison across all targets. Given $N_t$ numbers of input-output training pairs, the model was trained by minimizing the mean absolute percentage error (MAPE) loss function:
    \begin{align}
      \mathrm{MAPE} = \frac{100\%}{N_t}\sum_{i=1}^{N_t}
      \frac{| y^{(i)} - \hat{y}^{(i)}|}{| y^{(i)}|}. 
    \end{align}

Using Nadam as an optimizer \cite{Nadam}, performance is evaluated on $N_v$ validation dataset using Mean Absolute Error (MAE): 
  \begin{align}
\mathrm{MAE} = \frac{1}{N_v} \sum_{i=1}^{N_v} \bigl\lvert y^{(i)} - \hat{y}^{(i)} \bigr\rvert. 
  \end{align}

\section{Results and Discussion}\label{Sec:results}
\begin{figure}[ht]
\centering
\includegraphics[width=0.8\linewidth, frame]{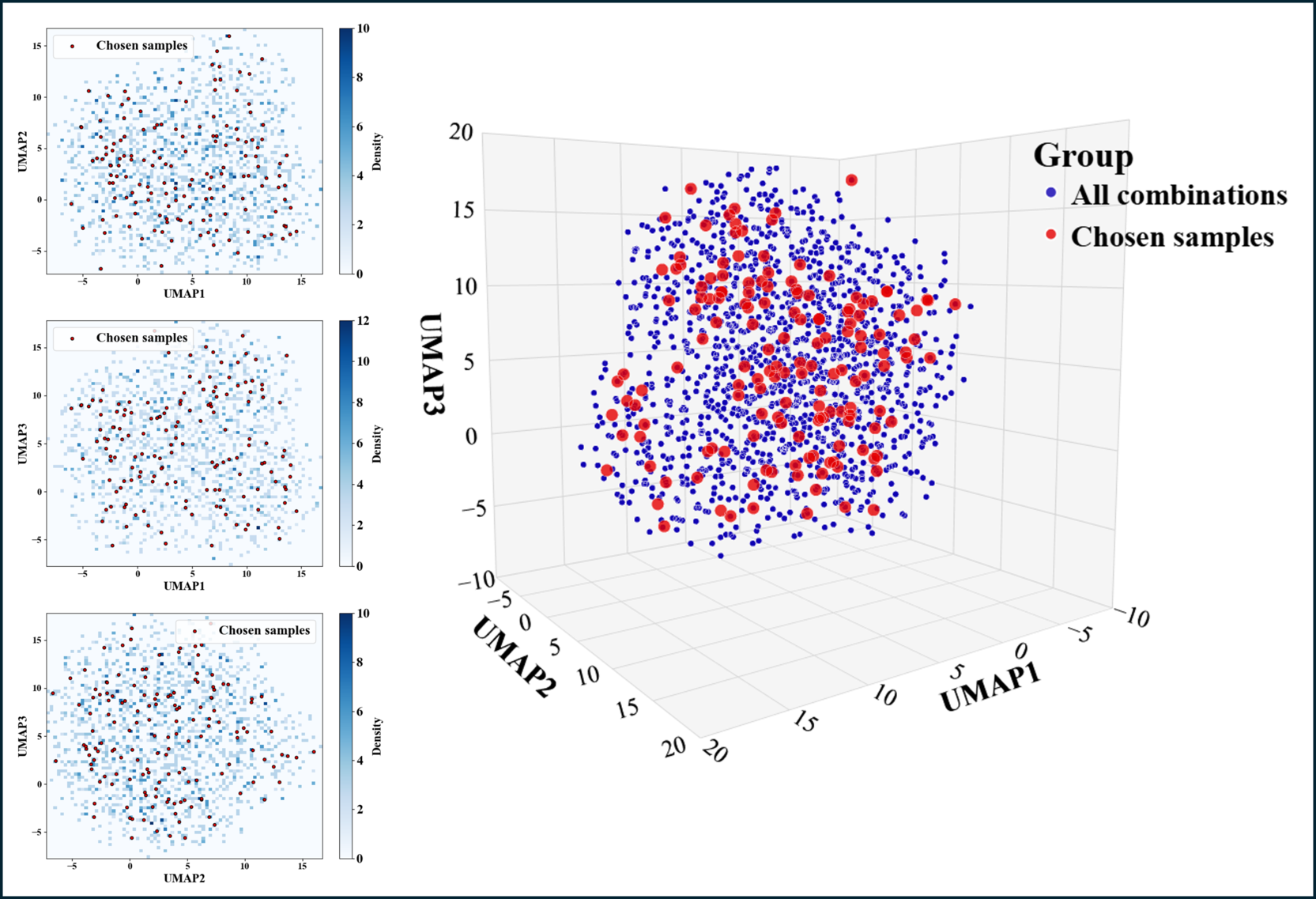}
\caption{UMAP-based visualization of the CFRC-AM design space and the LHS-selected experimental subset. \textbf{Right:} a 3D UMAP embedding showing all $4{,}320$ possible parameter combinations (blue) together with the $155$ chosen specimens (red). \textbf{Left:} the corresponding 2D projections, UMAP1 vs.\ UMAP2 (top), UMAP1 vs.\ UMAP3 (middle), and UMAP2 vs.\ UMAP3 (bottom) with density heat maps for all combinations; red markers indicate the selected subset.}
\label{fig:UMAP}
\end{figure}

We propose SE-WDNN to predict the economic and mechanical properties of CFRC-AM specimens. In this study, all computational experiments were executed in Python 3.10 on a workstation configured with an Intel Xeon W5-3435X processor (18 cores, 36 threads, 3.00 GHz) and an NVIDIA RTX 4500 Ada Generation GPU equipped with 24 GB of VRAM. 
To construct the dataset, the LHS method was used to select $155$ specimens out of $4,320$ possible samples, representing approximately $3.6\%$ of all combinations. Uniform Manifold Approximation and Projection (UMAP) \cite{mcinnes2018umap} was applied to reduce our seven-dimensional feature space into a three-dimensional embedding for better visualization, as illustrated in Figure \ref{fig:UMAP}. The uniform dispersion of the selected samples (red points) across the feature space confirms that the sampling method provides broad and unbiased coverage. 

During prediction, several supervised ML algorithms including feedforward neural network (FFN), Kolmogorov-Arnold network (KAN), XGBoost, CatBoost, and RF were benchmarked against SE-WDNN. 

\texttt{Optuna}, an open-source framework for automated hyperparameter optimization \cite{optuna_2019}, was used to fine-tune the models. The optimized hyperparameters and their corresponding search range for each model are summarized in Table \ref {tab:tuned hyperparameter}.

\begin{table*}[ht]
\centering
\caption{Model architectures and tuned hyperparameters (with ranges explored during search).}\label{tab:tuned hyperparameter}
\resizebox{1\linewidth}{!}{
\begin{tabular}{|l| c| c| c|}
\hline
\textbf{Models} & \textbf{Hyperparameters} & \textbf{Value} & \textbf{Range studied} \\
\hline \hline
\multirow{5}{2.3cm}{FFN} 
 & Layer architecture & (64, 16) & 1--3 layers; $[8\!:\!8\!:\!128]$\\
 & Activation Function & ReLU & [ReLU, Tanh, SELU, ELU]\\
 & Dropout Rate & 0.1 & 0.0--0.5 \\
 & Learning Rate & 0.01 & $10^{-4}$--$10^{-2}$ \\
 & Optimizer & RMSprop & [Adam, RMSprop, SGD, Adagrad, Adamax] \\
\hline

\multirow{4}{2.3cm}{CatBoost} 
 & Iterations (trees) & 1200 & [200, 800, 1200] \\
 & Depth (tree depth) & 4 & [4, 6, 8, 10, 12] \\
 & Learning Rate & 0.05 & [0.001, 0.01, 0.05, 0.1, 0.2] \\
 & Verbose / Random State & 0 \,/\, 25 & Fixed \\
\hline

\multirow{7}{2.3cm}{KAN} 
 & Architecture (width) & $[13,40,40,8]$ & Hidden layers: 1--2; $H\in\{27,\,40\}$ \\
 & Grid Resolution ($\texttt{grid}$) & 7 & [3, 5, 7] \\
 & Basis Order ($\texttt{k}$) & 2 & [2, 3, 4] \\
 & Optimizer & LBFGS & [LBFGS, Adam] \\
 & Learning Rate & $10^{-3}$ & [$10^{-2}$, $10^{-3}$, $10^{-4}$] \\
 & Training Steps & 80 & [80, 200, 1000] \\
 & Regularization ($\lambda$) & 0.0 & [0.0, $10^{-4}$, $10^{-3}$] \\
\hline

\multirow{4}{2.3cm}{XGBoost} 
 & Estimators (trees) & 100 & [100, 300, 500, 800, 1200] \\
 & Max Depth & 5 & [3, 5, 7, 9, 11] \\
 & Learning Rate & 0.05 & [0.001, 0.01, 0.05, 0.1, 0.2] \\
 & Subsample & 0.5 & [0.5, 0.7, 0.9, 1.0] \\
\hline

\multirow{4}{2.3cm}{RF} 
 & Estimators (trees) & 200 & [100, 200, 300, 500] \\
 & Max Depth & 10 & [None, 10, 20, 30] \\
 & Min Samples Split & 2 & [2, 5, 10] \\
 & Min Samples Leaf & 1 & [1, 2, 4] \\
\hline
\multirow{7}{2.3cm}{SE-WDNN} 
 & Deep Model Architecture & (128, 64) & 1--3 layers; $[8\!:\!8\!:\!128]$\\
 & Activation Function & ReLU & [ReLU, Tanh, SELU, ELU]\\
 & Dropout Rate & 0.1 & 0.0--0.5 \\
 & SE Block & (32(ReLU), 64(Sigmoid)) & 0.0--0.5 \\
 & Optimizer & Nadam & [Adam, RMSprop, Nadam, Adagrad, AdamW] \\
 & Learning Rate & 0.01 & $10^{-4}$--$10^{-2}$ \\
 & Batch Size & 16 & 2--32 \\
\hline
\end{tabular}
}
\end{table*}

\noindent\textbf{FFN} was implemented with two hidden layers containing $64$ and $16$ neurons, respectively, and used the Rectified Linear Unit (ReLU) activation function. Each hidden layer included dropout regularization with a rate of $0.1$ to reduce overfitting. 
Training used the RMSprop optimizer \cite{Rmsprop} with a learning rate of $0.01$ and MAPE as the loss function. Early stopping with patience of $10$ epochs was applied to further reduce the risk of overfitting.

\noindent\textbf{KAN} is a recently introduced neural architecture that replaces conventional linear weights with learnable nonlinear one-dimensional functions, providing greater interpretability and expressive power for high‑dimensional function approximation \cite{liu2024kan}. In this study, the network consisted of two hidden layers, each with a width of 22, allowing complex mappings through compositions of univariate functions. The model was trained with the Limited-memory Broyden-Fletcher-Goldfarb-Shanno (L-BFGS) \cite{L-bfgs} optimizer for 80 iterations at a learning rate of 0.01, without regularization or added noise. To control the level of functional detail, a grid resolution of 7 and a basis complexity of 2 were applied.

\noindent\textbf{XGBoost} is a well-known gradient-boosted decision tree with the ability to capture complex, nonlinear relationships \cite{XGboost}. 
In this study, the base regressor was configured with $100$ estimators, a learning rate of $0.05$, a maximum tree depth of 5, and a sub-sampling ratio of 0.5 to balance bias and variance.

\noindent\textbf{CatBoost} is a gradient boosting algorithm designed with built‑in optimizations for handling categorical data, making it well‑suited for multi‑output regression \cite{dorogush2018catboost}. Unlike traditional approaches that require explicit one‑hot encoding, CatBoost processes categorical features internally using ordered boosting and target statistics, which helps reduce overfitting and improve generalization. In this study, the base CatBoost regressor was configured with 1200 iterations, a learning rate of 0.05, and a tree depth of 4.

\noindent \textbf{RF} regressor is a group of decision trees recognized for its strength and ability to avoid overfitting. We ran the model with 200 trees and a maximum depth of 10, while keeping the other parameters in their default settings.

\subsection{Model Performance Analysis}
\begin{figure}[ht]
 \centering
 \includegraphics[width=0.9\linewidth, frame]{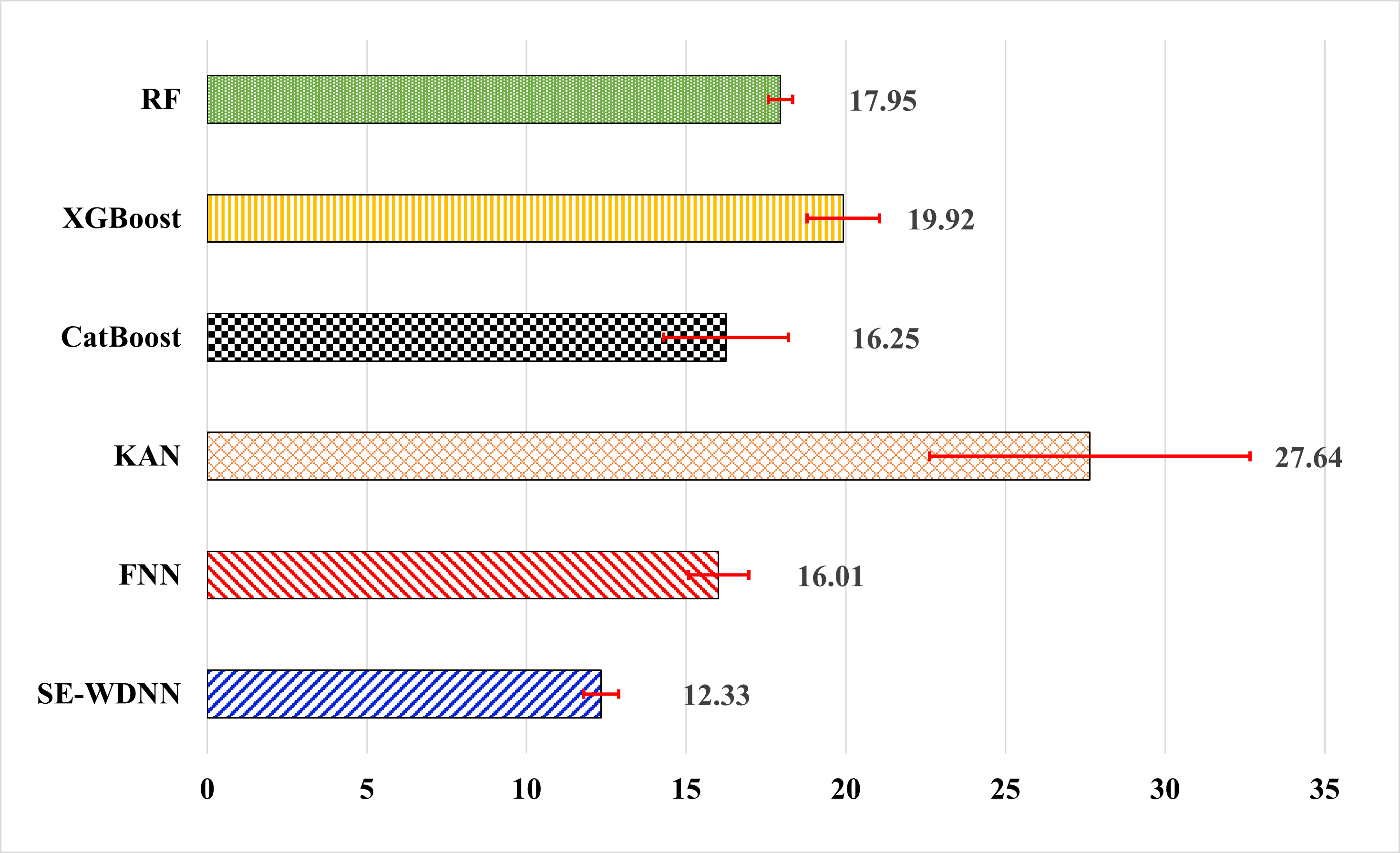}
 \caption{Comparison of overall predictive performance (MAPE\%) across six ML models for estimating mechanical and manufacturing properties of fiber-reinforced polymer composites. The Wide and Deep Neural Network with Squeeze-and-Excitation achieves the lowest error, while KAN shows the highest variability and overall error. Error bars represent standard deviation across five runs.}
 \label{fig:Overall MAPE}
\end{figure}

The comparative performance of all six models, calculated from predictions and actual values on the unseen test dataset, is illustrated in Figure \ref{fig:Overall MAPE}. For all models, a random split assigned $80\%$ of the data to the training set and the other $20\%$ to the test set. This method allows us to train models on a larger portion of the data and evaluate their performance and prediction accuracy on an unseen dataset to assess how well the model performs on new scenarios. To account for randomness, each model was trained five times with different random initializations while keeping the split fixed. The values reported in Figure \ref{fig:Overall MAPE} are the mean test-set metrics from these five runs, and the error bars display the standard deviation calculated over the five seeds. 

The SE‑WDNN model achieved the lowest MAPE ($12.33\%$) among all six models, demonstrating the highest prediction accuracy. Incorporating SE blocks into WDNN improved the model's ability to capture both low‑level feature interactions (through the wide component) and higher‑order nonlinear relationships (through the deep component). The FFN and CatBoost models showed nearly the same performance, with MAPE values of $16.25\%$ and $16.01\%$ respectively. CatBoost's impressive results stem from its ability to manage categorical variables without much preprocessing, its use of the ordered boosting method, and its reduced need for hyperparameter tuning. RF and XGBoost produced moderate results, with MAPE values of $17.95\%$ and $19.92\%$, respectively, indicating that they may lack the architectural flexibility needed to capture the complex interaction patterns in this dataset fully. KAN delivered the weakest performance, recording the highest MAPE ($27.64\%$) and the largest standard deviation across trials. These results indicate that, despite KAN's theoretical strengths in function approximation, it may be prone to instability or overfitting when applied to mixed‑type structured datasets of limited size.

\begin{figure*}[!ht]
\centering
\includegraphics[width=0.9\linewidth, frame]{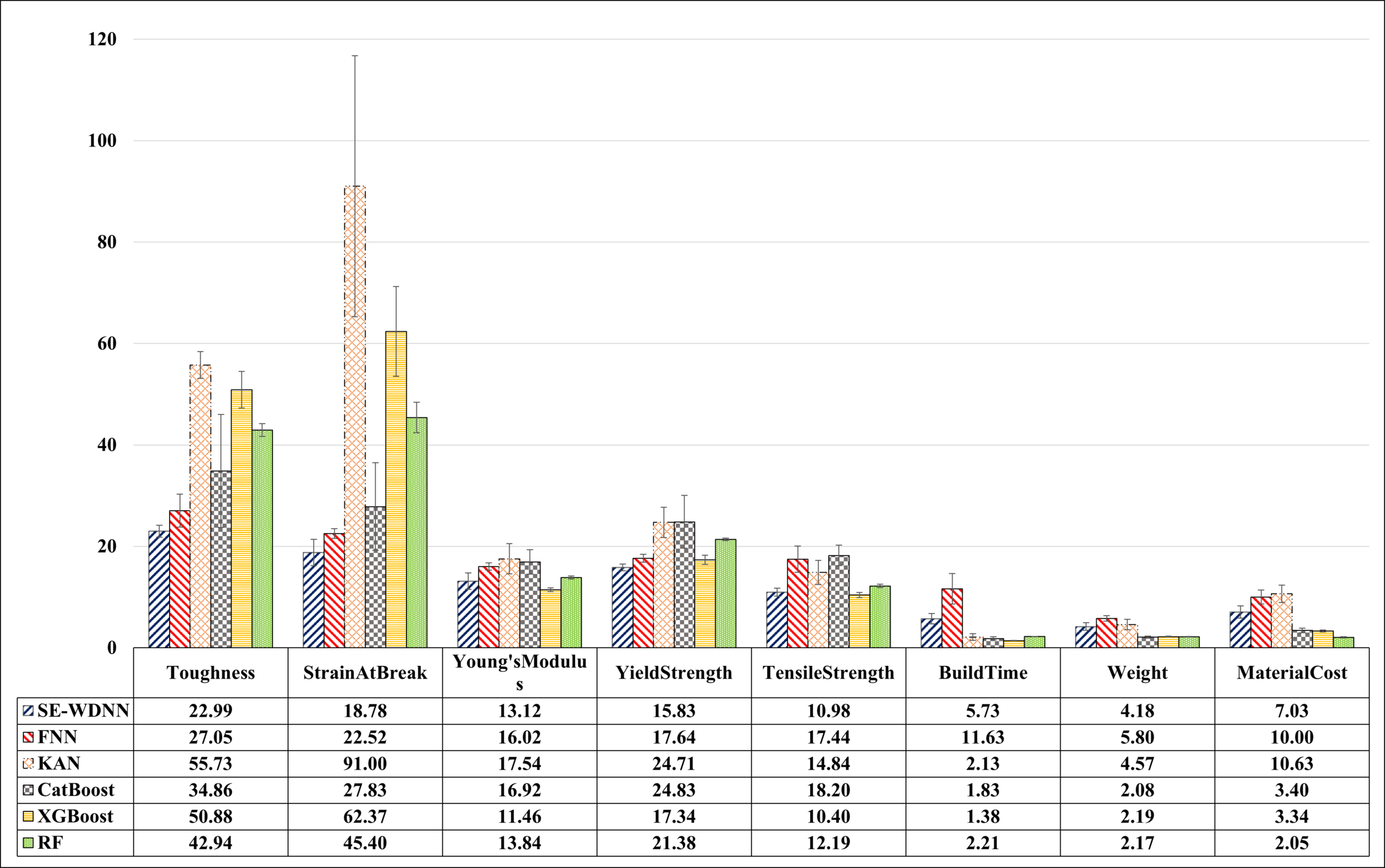}
\caption{Models' predictive performance across eight target properties of fiber-reinforced polymer composites was evaluated using $MAPE\%$.}
\label{fig:MAPE-Target}
\end{figure*}

To examine the predictive behavior of the six ML models across different target variables, we analyzed their MAPE on each of the eight targets, as shown in the bar charts in Figure \ref{fig:MAPE-Target}. 

\paragraph{\textbf{Toughness}} 
The SE‑WDNN achieved the lowest error ( MAPE = 22.99\% ), outperforming the FFN (27.05\%) and all tree‑based models. In contrast, KAN (55.73\%) and XGBoost (50.88\%) performed poorly, likely because the relatively small magnitude of this variable amplifies percentage errors. CatBoost also exhibited high variability (Std Dev = 11.13\%), indicating instability across samples.

\paragraph{\textbf{Strain at Break}} 
Once again, the SE‑WDNN delivered the lowest error (18.78\%), while KAN recorded the worst performance (91.00\%, Std Dev=35.74\%), highlighting its sensitivity to the extreme outlier in this highly skewed target. The FFN and CatBoost models showed moderate performance, whereas RF and XGBoost achieved MAPE values above 45\%.

\paragraph{\textbf{Young's Modulus}}
XGBoost delivered the most accurate predictions (MAPE=11.46\%), followed closely by SE‑WDNN (13.12\%) and RF (13.84\%). KAN and FFN performed slightly less effectively, while CatBoost provided balanced performance with moderate variance.

\paragraph{\textbf{Yield Strength}} 
The SE‑WDNN and FFN achieved the best results, with MAPE values of 15.83\% and 17.64\%, respectively. CatBoost and KAN showed higher errors and greater variability, while RF delivered consistent but less accurate predictions (21.38\%).

\paragraph{\textbf{Tensile Strength}} 
XGBoost recorded the lowest error (10.40\%), with SE‑WDNN (10.98\%) and RF (12.19\%) close behind. Interestingly, KAN performed comparatively well (14.84\%) relative to its results on other targets, likely benefiting from the more predictable relationship between the input features and tensile strength.

\paragraph{\textbf{Build Time}} 
All models performed well on this target, with XGBoost (1.38\%) and CatBoost (1.83\%) achieving the lowest errors. KAN also delivered strong results (2.13\%), likely aided by the target's limited range. SE‑WDNN posted a slightly higher error (5.73\%).

\paragraph{\textbf{Weight}} 
CatBoost (2.08\%), XGBoost (2.19\%), and RF (2.17\%) emerged as the top performers. FFN produced the highest error (5.80\%), indicating that tree‑based methods may be better suited for predicting this low‑variance variable.

\paragraph{\textbf{Material Cost}} 
RF (2.05\%) produced the most accurate and stable predictions, followed by CatBoost (3.40\%) and XGBoost (3.34\%). Neural networks, especially FFN and KAN, yielded significantly higher errors (MAPE > 10\%).

To better understand the higher prediction error of strain at break, we analyzed the measurement variability using only the 35 replicated processing conditions (70 tensile tests). For each property, the coefficient of variation (CV) was computed from the two replicate measurements as
\[
\mathrm{CV} = \frac{\sigma}{\mu} \times 100\%,
\]
where $\sigma$ is the sample standard deviation and $\mu$ is the corresponding mean of the two replicate measurements. Although the median CV for strain at break and the strength-related parameters are similar(3-4\%), strain at break shows much greater extreme variability. Its CV values can reach up to 26 percent, while the maximum CVs for strength and modulus are only 13 to 16 percent. In other words, strain at break has occasional large fluctuations that strength and stiffness do not show. This wider and less regular variability makes strain at break harder to model, which explains its higher prediction error compared to other mechanical properties.

\begin{figure}[!ht]
 \centering
 \includegraphics[width=0.7\linewidth]{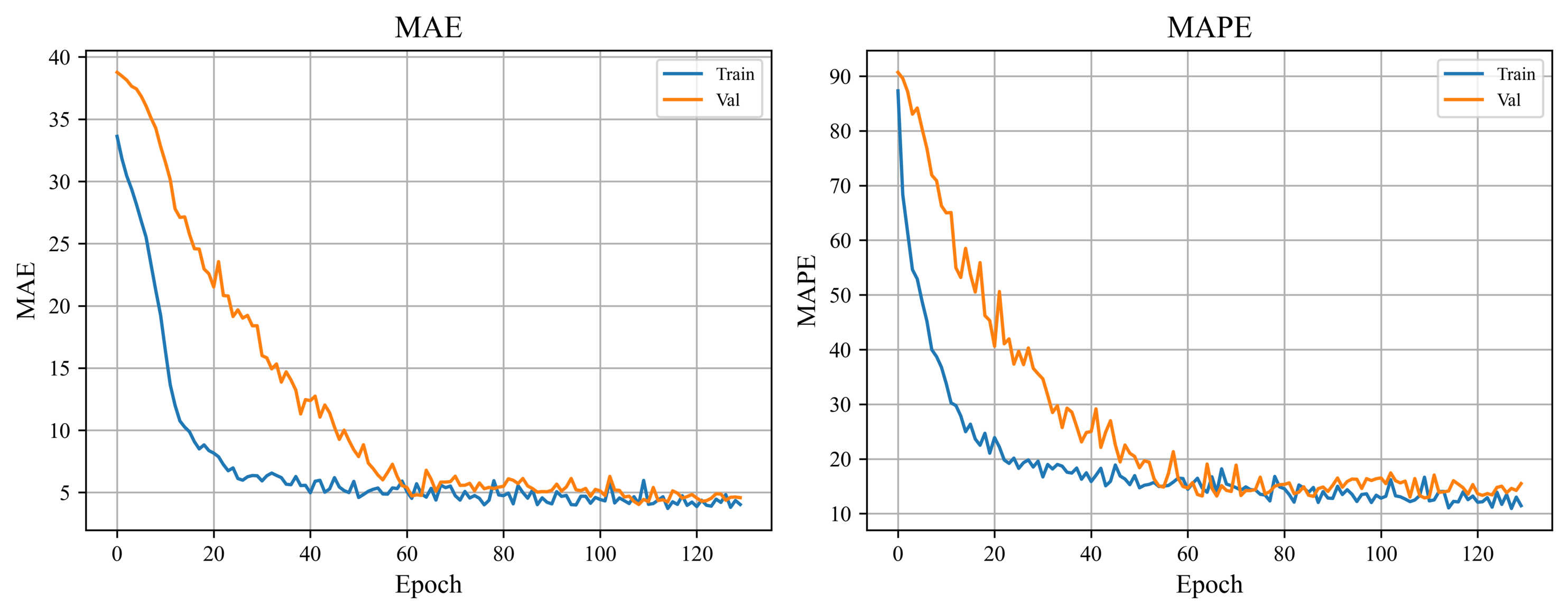}
 \caption{Training and validation curves of the SE-WDNN model over 125 epochs based on MAE and MAPE.}
 \label{fig:validation curve}
\end{figure}

Figure \ref{fig:validation curve} shows the training and validation trends of the SE‑WDNN model over $125$ epochs, evaluated using both MAE and MAPE. 
In both metrics, the curves indicate effective learning and convergence. The validation loss closely tracks the training loss without notable divergence, suggesting minimal overfitting and strong generalization. Around the 60\textsuperscript{th} epoch mark, the validation performance stabilizes, with MAE and MAPE maintaining low values that remain nearly parallel to the training curves. Overall, these patterns highlight the robustness of the SE‑WDNN architecture in capturing complex nonlinear relationships within the dataset while avoiding excessive variance.

\begin{figure*}[!ht]
 \centering
 \includegraphics[width=1\linewidth, frame]{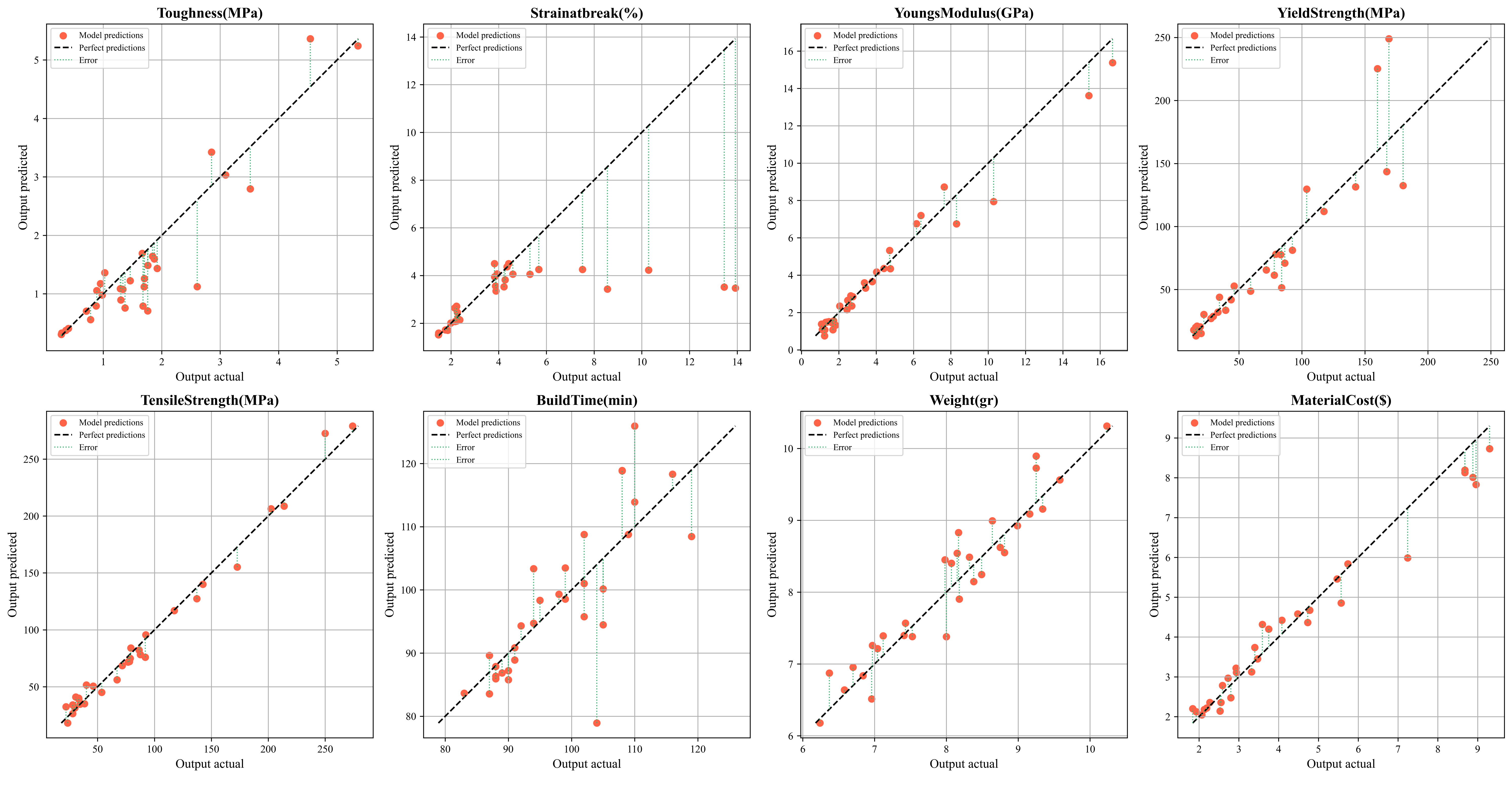}
 \caption{Predicted versus actual values for all eight target variables using the SE-WDNN model. Each subplot displays the model's predicted output (y-axis) against the ground truth (x-axis), with the dashed black line representing perfect predictions and green dotted lines indicating prediction errors.}
 \label{fig:Predicted-Actual}
\end{figure*}

Figure \ref{fig:Predicted-Actual} presents scatter plots of predicted versus actual values for each of the eight output variables generated by the SE‑WDNN model. Overall, the plots reveal strong agreement between predictions and actual measurements, particularly for Tensile Strength, Young's Modulus, and Material Cost, highlighting the model's high predictive accuracy and generalization capability. In contrast, somewhat larger deviations are noticeable for targets such as Strain at Break and Yield Strength.

\section{Ablation Study} \label{sec:ablation}
\subsection{SHAP Analysis}
\begin{figure*}[!t]
    \centering
    \begin{subfigure}[b]{0.3\textwidth}
        \includegraphics[width=\linewidth]{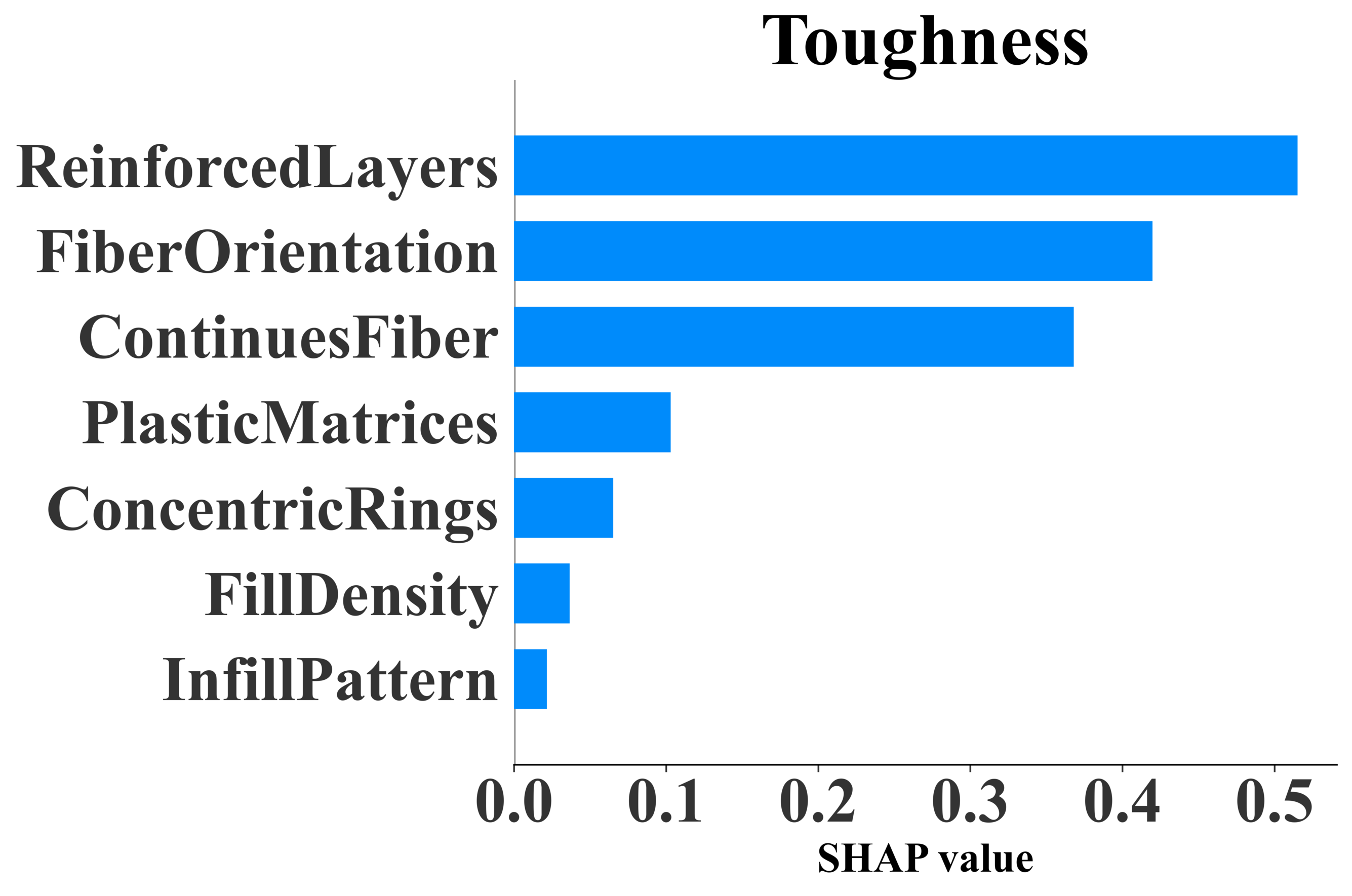}
    \end{subfigure}
    \begin{subfigure}[b]{0.3\textwidth}
        \includegraphics[width=\linewidth]{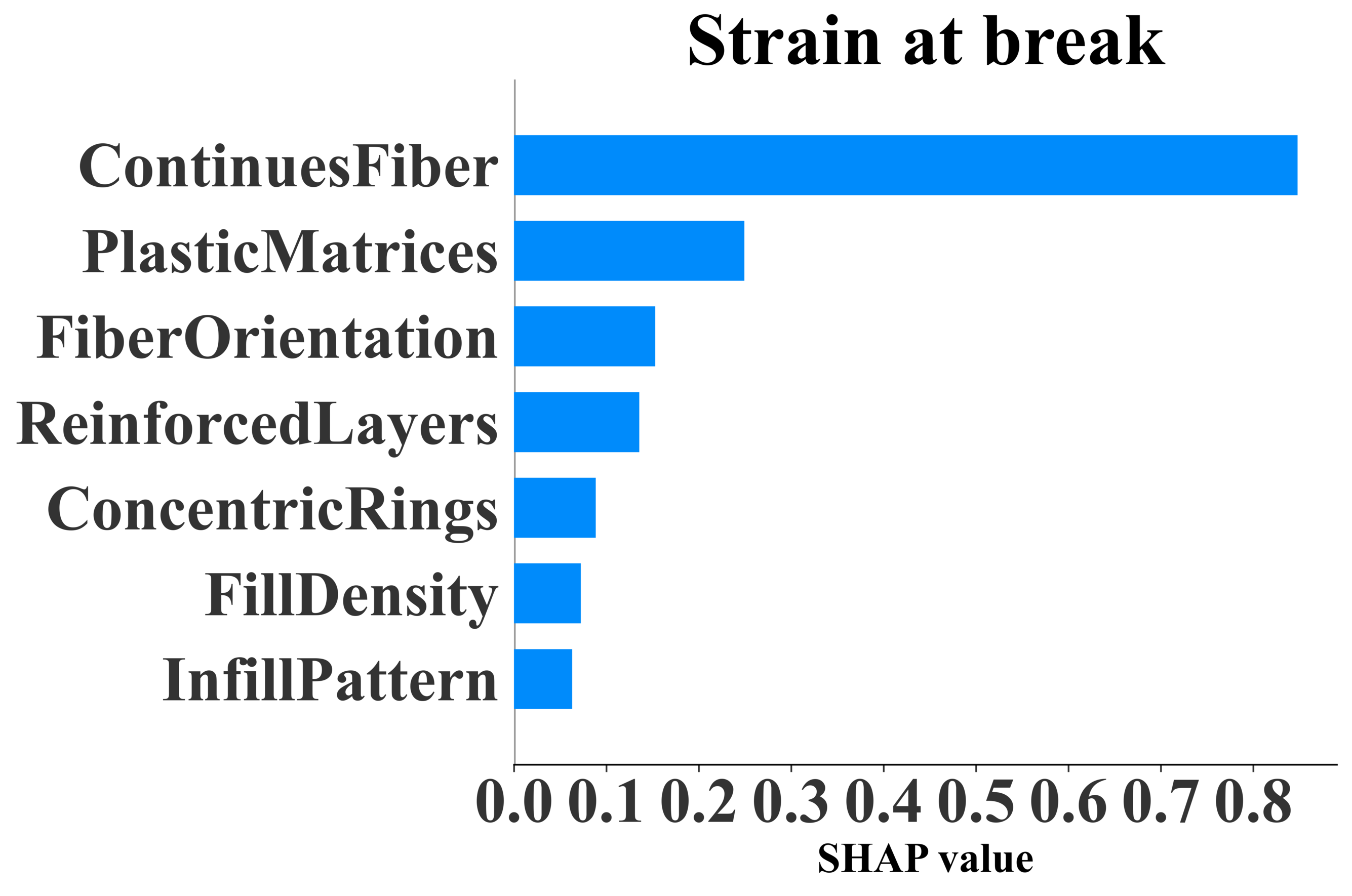}
    \end{subfigure}
    \begin{subfigure}[b]{0.3\textwidth}
        \includegraphics[width=\linewidth]{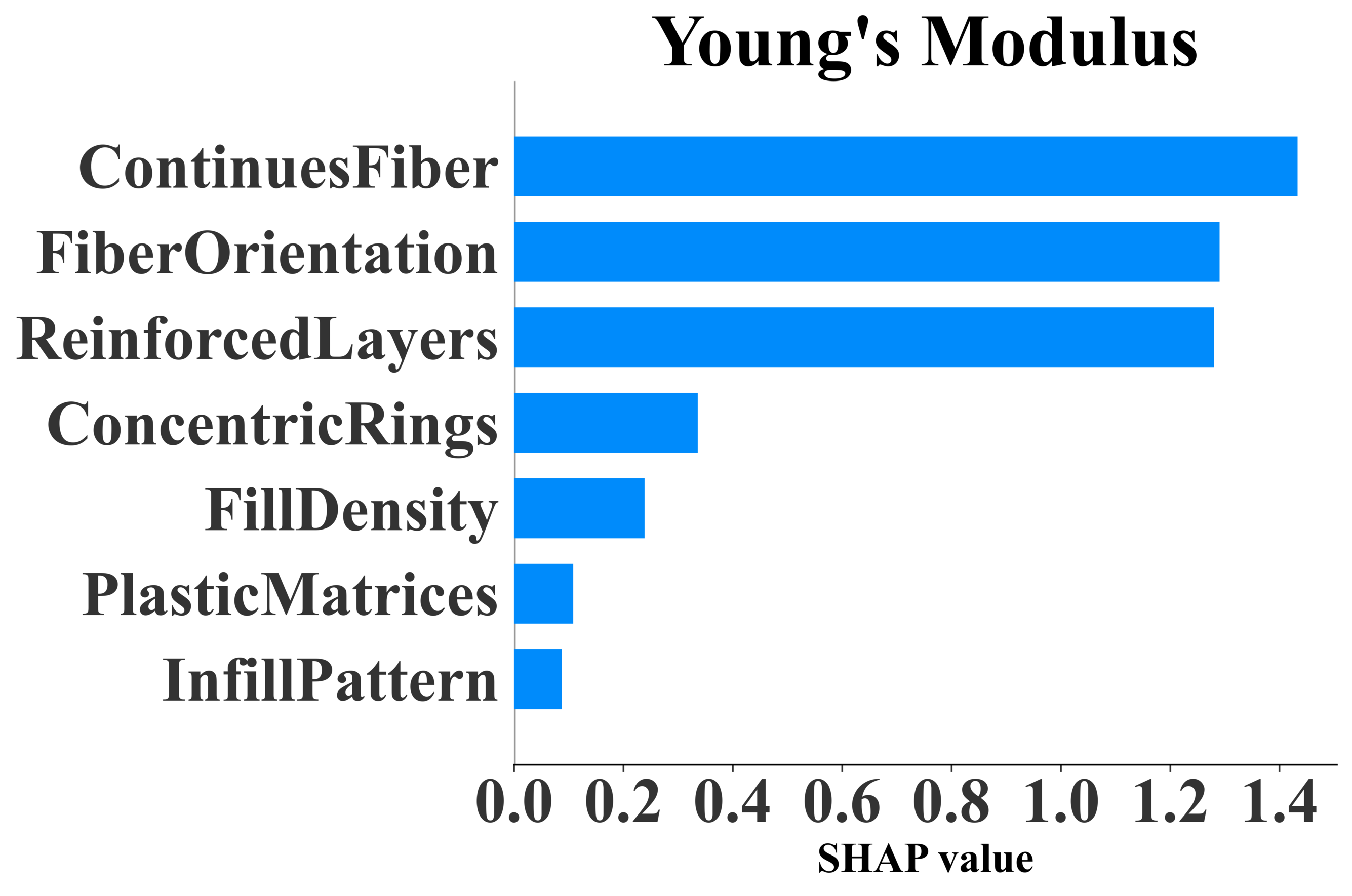}
    \end{subfigure}

    \begin{subfigure}[b]{0.3\textwidth}
        \includegraphics[width=\linewidth]{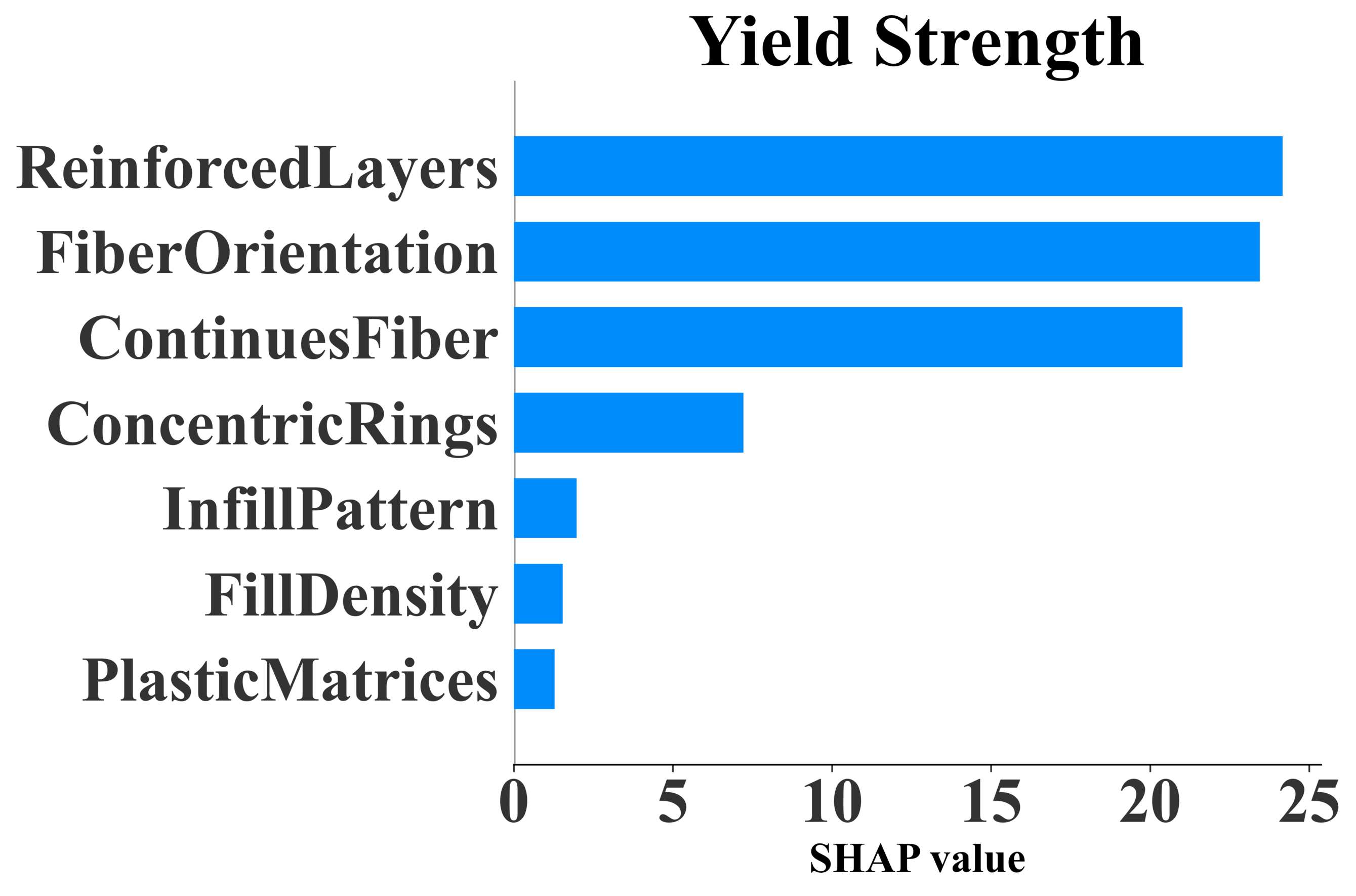}
    \end{subfigure}
    \begin{subfigure}[b]{0.3\textwidth}
        \includegraphics[width=\linewidth]{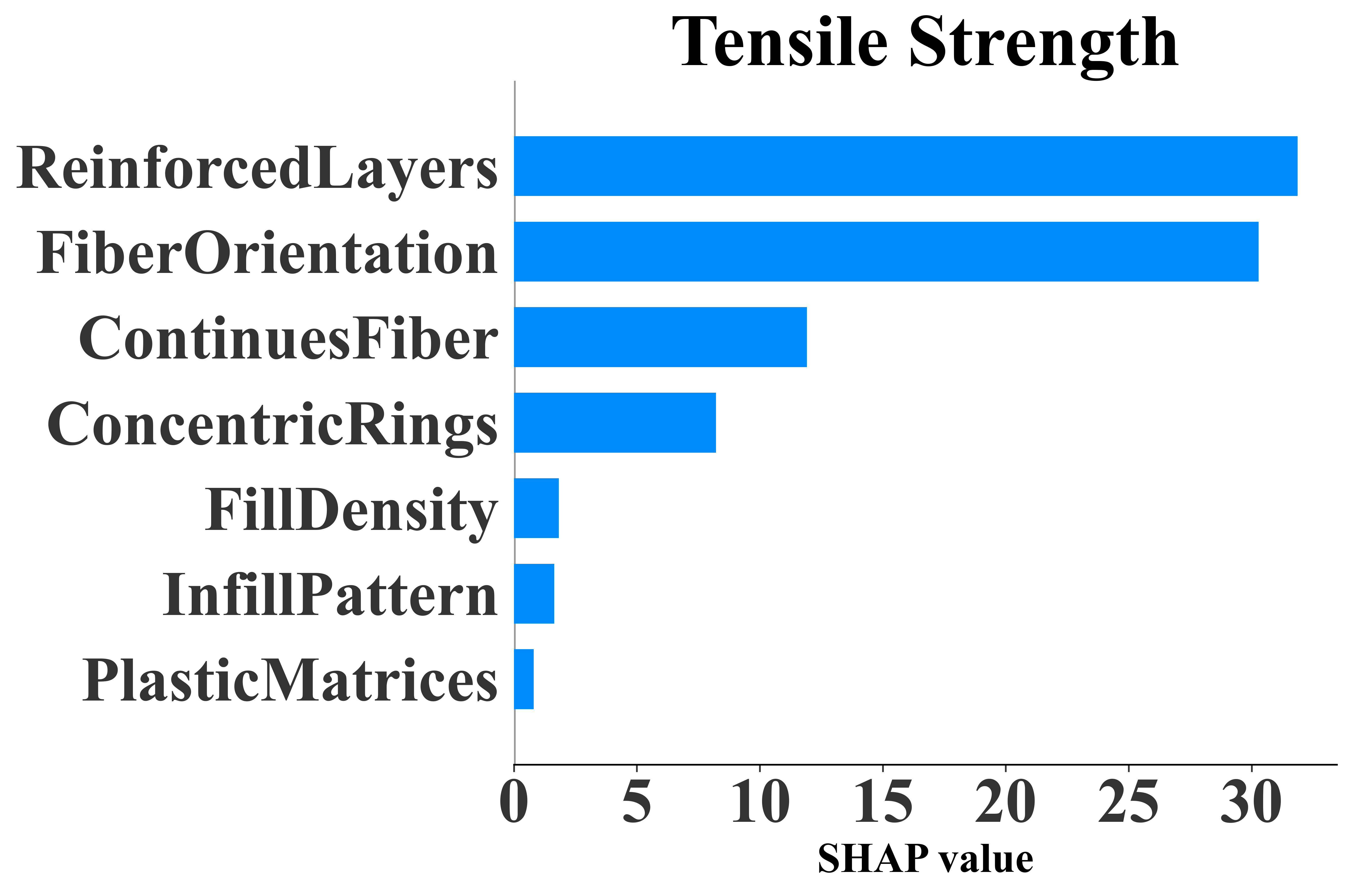}
    \end{subfigure}
    \begin{subfigure}[b]{0.3\textwidth}
        \includegraphics[width=\linewidth]{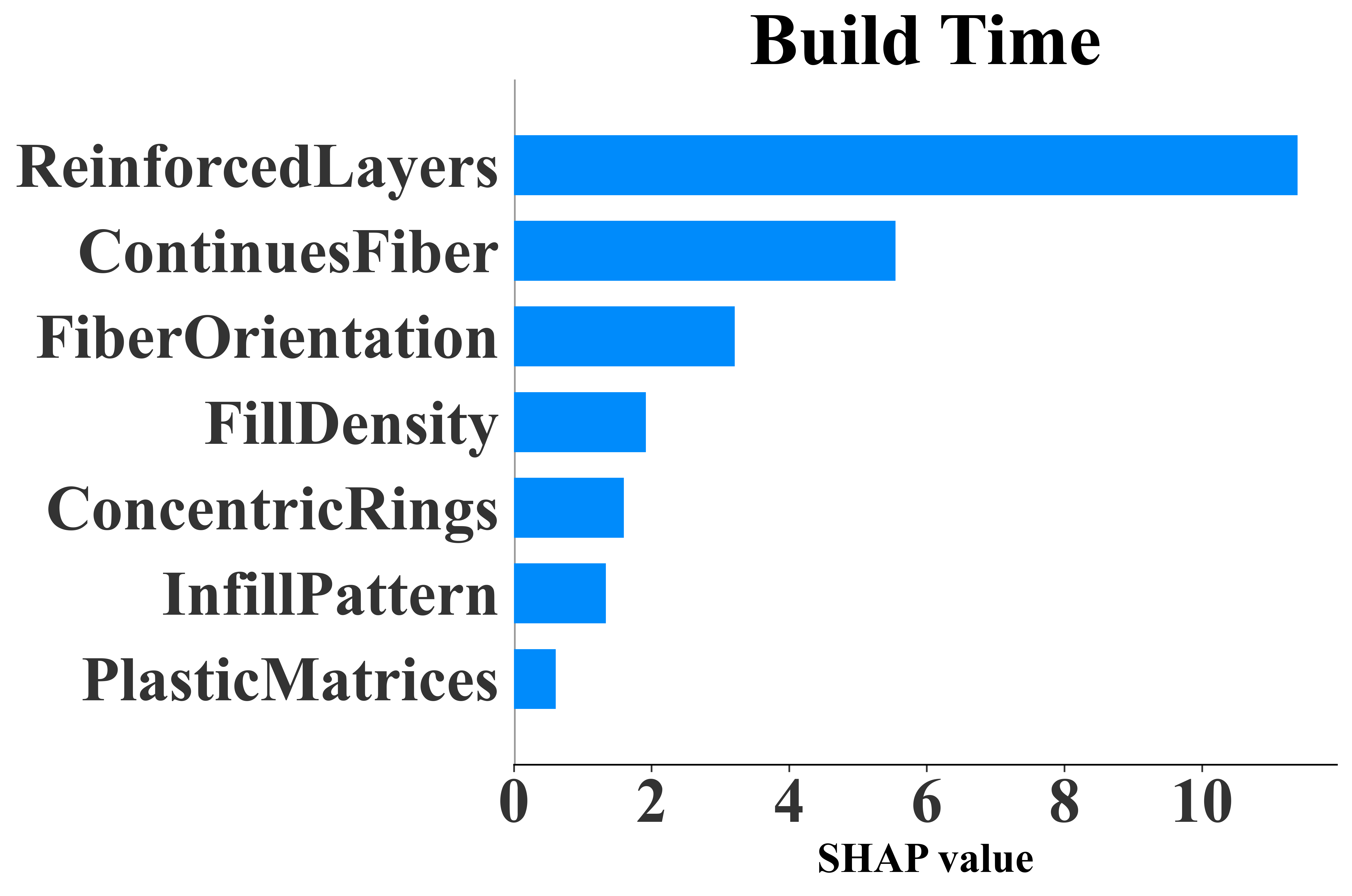}
    \end{subfigure}

    \begin{subfigure}[b]{0.3\textwidth}
        \includegraphics[width=\linewidth]{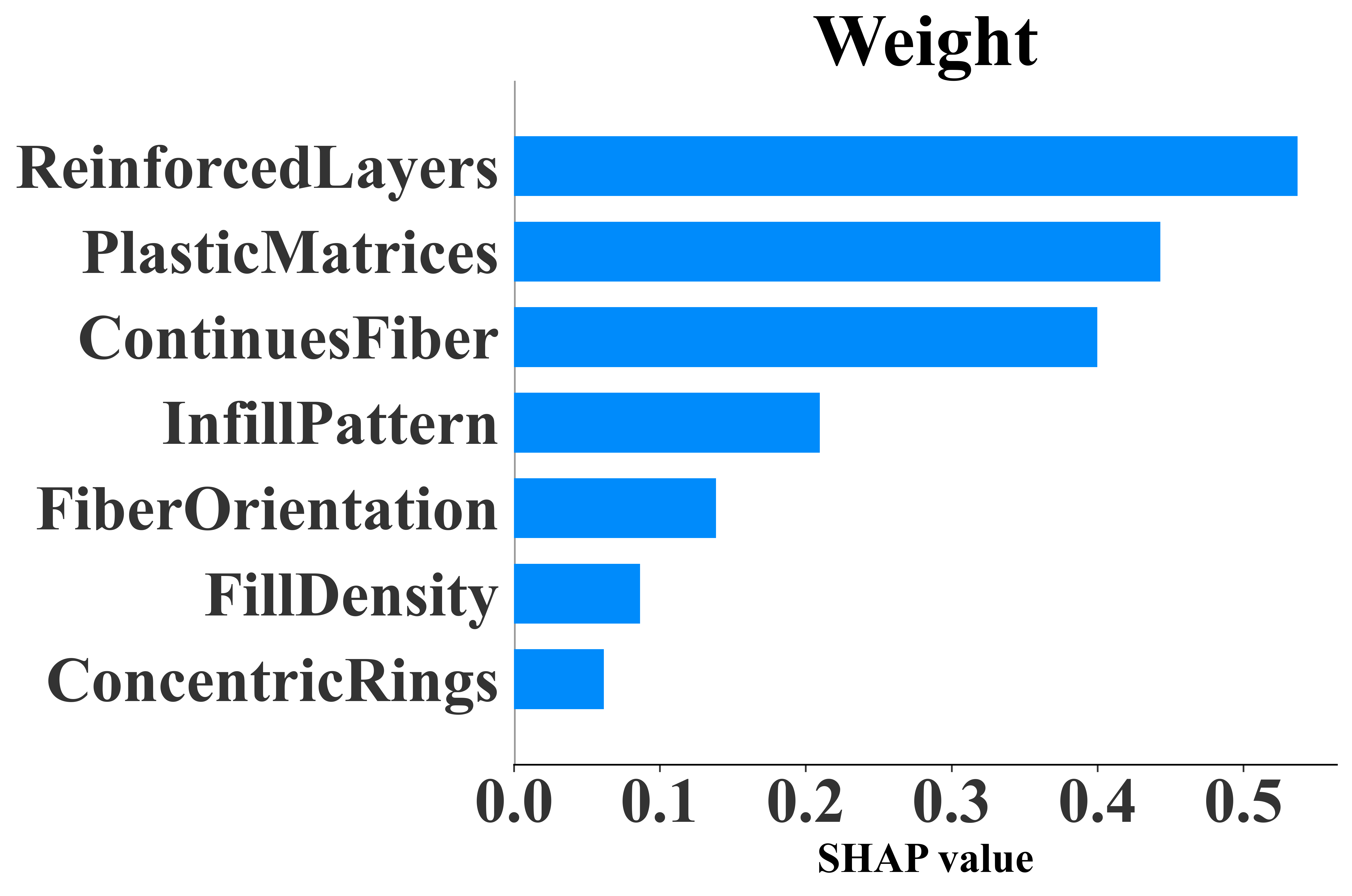}
    \end{subfigure}
    \begin{subfigure}[b]{0.3\textwidth}
        \includegraphics[width=\linewidth]{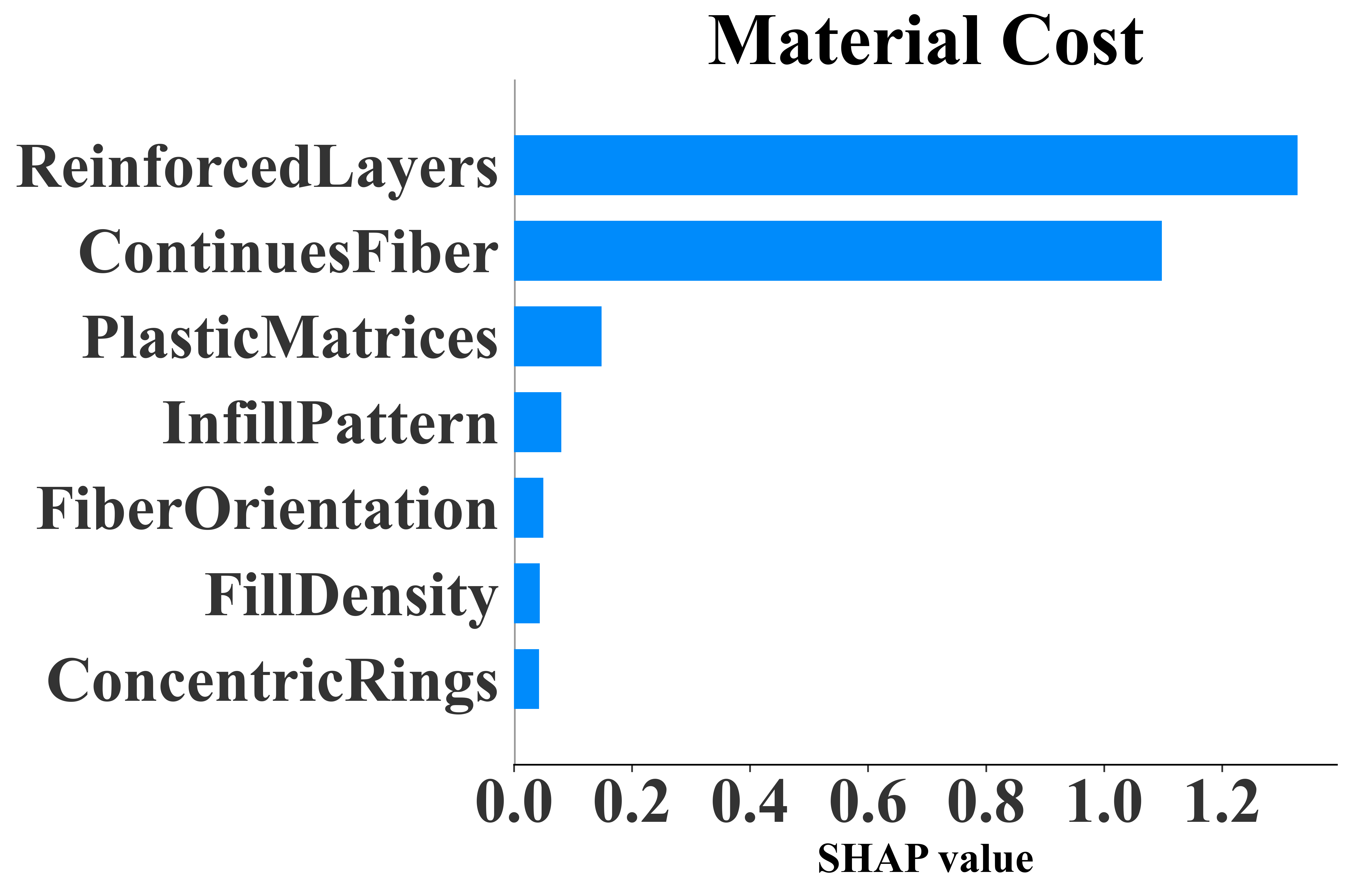}
    \end{subfigure}
    \begin{subfigure}[b]{0.3\textwidth}
    \end{subfigure}
    
    \caption{SHAP summary plots for all target variables in the composite dataset. Each bar depicts the mean absolute SHAP value of a feature, reflecting its overall contribution to the model's predictions for a given target. Reinforcement-related parameters such as reinforced layers, continuous fiber (fiber type), and fiber orientation consistently rank highest for mechanical targets, underscoring their dominant influence on composite behavior.}
    \label{fig:Shap}
\end{figure*}

To better understand the inner workings of the SE-WDNN model, SHAP was applied to assess the relative importance of each input feature across all target outputs. As seen in Figure \ref{fig:Shap}, the analysis consistently highlighted three variables (the number of reinforced layers, the type of continuous fiber, and fiber orientation) as having the most significant influence on key mechanical properties such as toughness, Young's modulus, yield strength, and tensile strength. These features are intimately linked to the underlying reinforcement mechanisms in continuous fiber composites, reinforcing the notion that both the amount and type of reinforcing fiber, along with how those fibers are aligned within the structure, serve as primary drivers of mechanical performance.

When it comes to predicting strain at break, the type of continuous fiber emerged as the most impactful factor, emphasizing the critical role of material selection in shaping ductile behavior. Each of these fibers (Carbon Fiber, Fiberglass, and Kevlar) possesses unique mechanical characteristics, including stiffness, elongation capacity, and energy absorption, all of which contribute directly to how the material responds under strain before failure. In contrast, features such as type of plastic matrices, fill density, and infill pattern played a comparatively modest role across most mechanical targets, indicating their influence on structural performance is secondary. Shifting focus to production-related outcomes, Build Time was most affected by the number of reinforced layers and the type of continuous fiber. In contrast, weight was mainly determined by the number of reinforcing layers. The type of plastic matrices and the type of continuous fiber followed closely behind. Meanwhile, material cost was mostly influenced by factors related to reinforcement. The number of reinforced layers and the type of continuous fiber were the biggest contributors. Overall, the SHAP analysis makes it clear that composite performance is primarily driven by the reinforcement strategy, specifically, how the structure is configured (such as layer count and fiber orientation) and which materials are selected, particularly the type of fiber. These findings point to a key takeaway: optimizing both mechanical strength and cost efficiency hinges more on fiber-related choices than on matrix composition or geometric infill patterns. Recognizing this can meaningfully inform design and manufacturing strategies for CFRC-AMs, leading to more targeted and effective engineering decisions \cite{hasti}.
 \subsection{Understanding the Effect of SE Blocks on WDNN}
To understand how the SE block improves prediction accuracy, we first examined how it reweights the channels of the deep feature representation. The SE mechanism generates a 64-dimensional vector of attention weights for each sample. Each weight ranges from 0 to 1 and is multiplied by the corresponding channel. Larger weights increase the importance of key channels, while smaller weights lessen the effect of less informative ones. Looking at these attention patterns reveals which hidden features the SE block focuses on throughout the dataset.

\begin{figure*}[!ht]
 \centering
 \includegraphics[width=1\linewidth, frame]{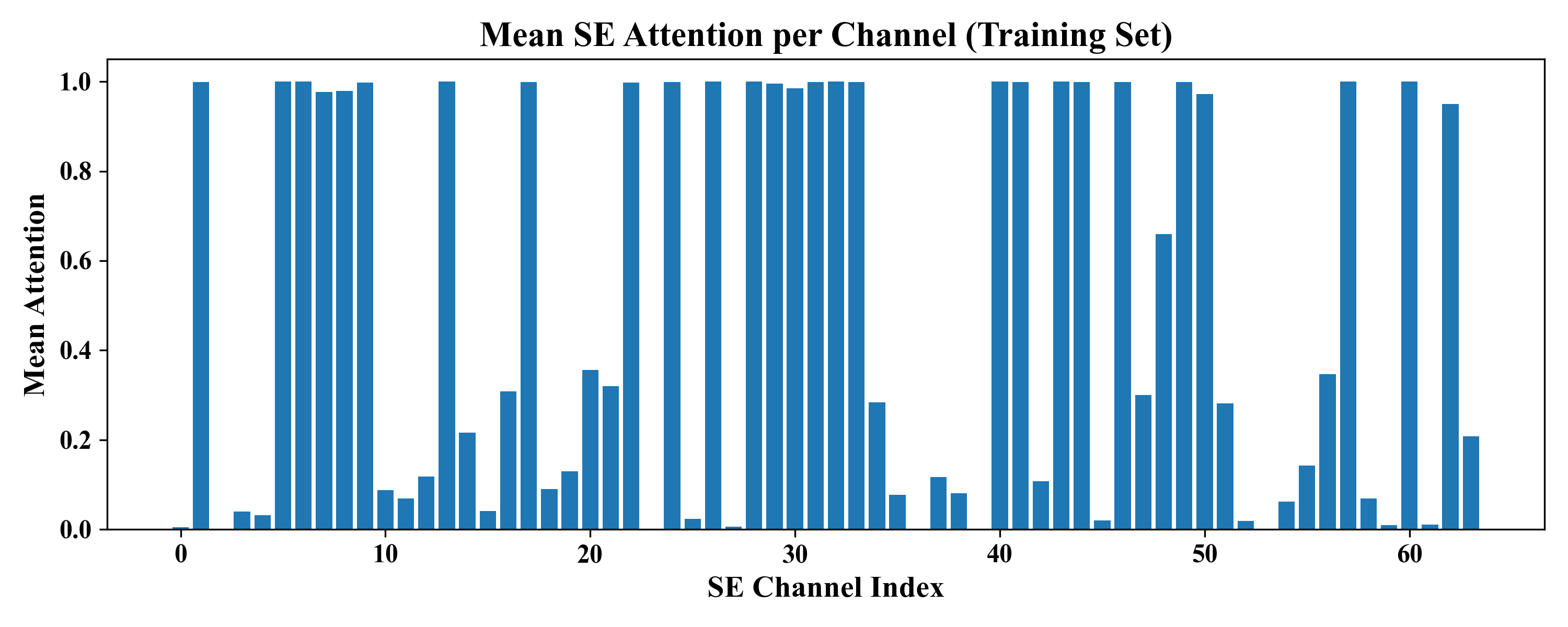}
 \caption{Mean SE weight per channel across all training samples. Only a few channels get high average weights. This shows that the SE block learns a small number of important latent features instead of evenly adjusting all channels.}
 \label{fig:se_mean}
\end{figure*}

The Figure \ref{fig:se_mean} shows that the SE block does not distribute attention evenly. It identifies a small number of channels that are consistently amplified across the training set, while most channels receive much lower weights. This selective focus matches the SHAP analysis, where parameters related to reinforcement stand out in importance. Overall, these findings suggest that the SE mechanism acts as a learned feature selector, improving the channels that respond best to the reinforcement strategy.

\begin{figure*}[!ht]
 \centering
 \includegraphics[width=1\linewidth, frame]{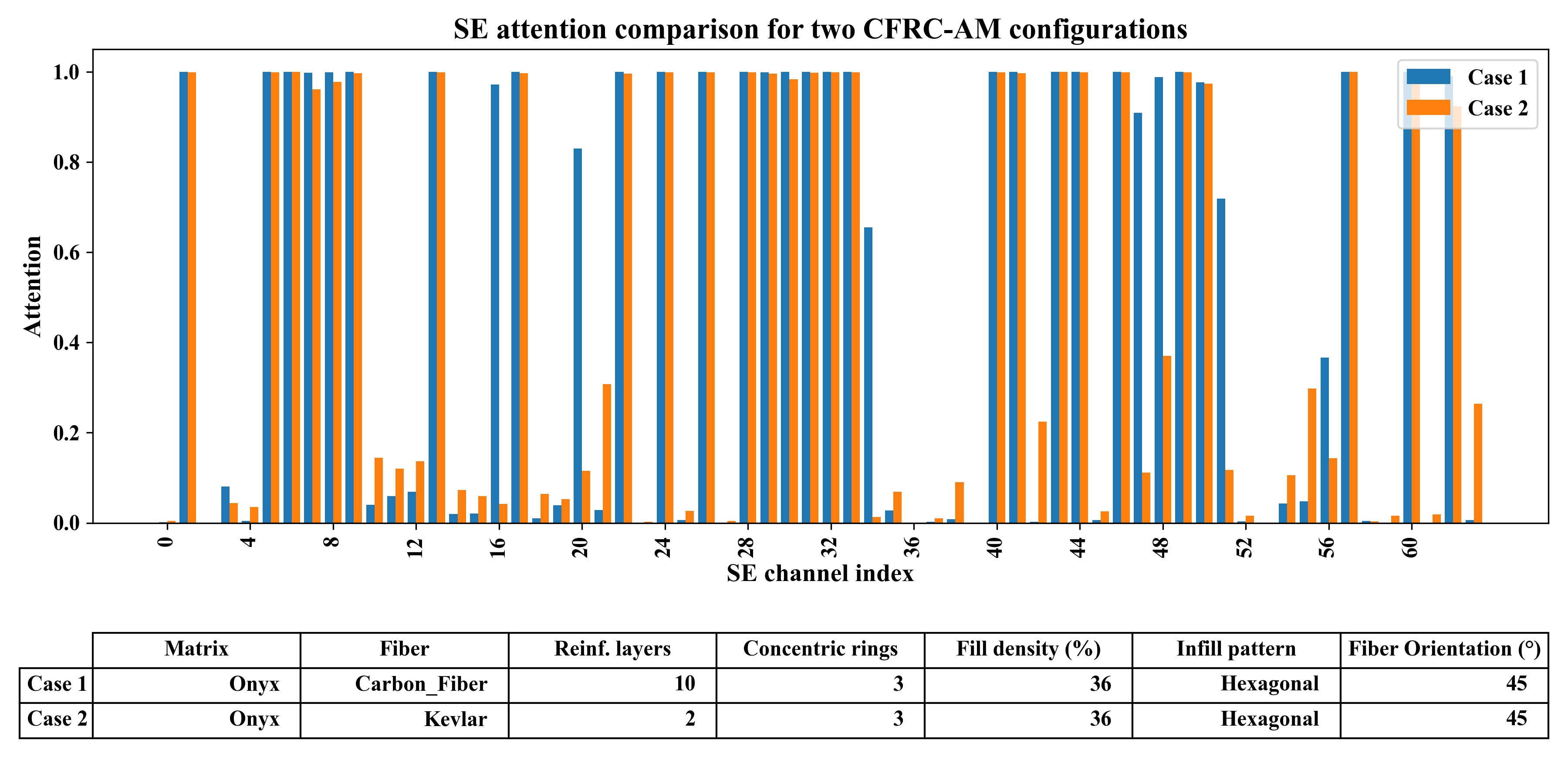}
 \caption{Comparison of SE attention vectors for two CFRC-AM configurations with the same infill and geometric parameters but different reinforcement strategies. Case 1 uses ten reinforced carbon-fiber layers. Case 2 uses two reinforced Kevlar layers. Several channels have much higher weights in the carbon-fiber case. This shows that the SE block is very sensitive to the type of reinforcement and the number of layers.}
 \label{fig:se_twocases}
\end{figure*}

To illustrate specimen-level behavior, Figure \ref{fig:se_twocases} compares the attention profiles of two configurations that differ only in reinforcement type and the number of reinforced layers. Because all geometric and infill parameters remain the same, differences in the attention vectors reflect the SE block's response to the reinforcement strategy. The carbon-fiber specimen with ten reinforced layers shows much stronger activation across several channels. In contrast, the Kevlar specimen with two layers creates a much sparser attention pattern.

Also, to understand the effect of including SE blocks in the WDNN model, we compute a paired t-test comparing baseline WDNN and WDNN with SE (SE-WDNN). The paired t-test was applied by first computing the MAPE difference for each target $j$ in every run:
\begin{align}
d^{(j)}_i = \mathrm{MAPE}^{(j)}_{\mathrm{WDNN},i} - \mathrm{MAPE}^{(j)}_{\mathrm{SE-WDNN},i}, \quad i = 1, 2, \dots, n,
\end{align}
\noindent where $n$ represents the number of runs. The mean and standard deviation of these differences are calculated as:
\begin{align}\label{eq:mst}
\bar{d} = \frac{1}{n} \sum_{i=1}^{n} d_i, \quad s_d = \sqrt{\frac{\sum_{i=1}^n (d_i - \bar{d})^2}{n-1}}.
\end{align}
The $t$-statistic is then computed from the results in Equation \ref{eq:mst} as:
\begin{align}
t = \frac{\bar{d}}{s_d / \sqrt{n}}.
\end{align}

The $p$-values were obtained from the Student's $t$-distribution. Table \ref{tab:paired_ttest_se} shows the results of a paired $t$-test comparing $n = 5$ independent runs of both models. For each target variable MAPE, the table includes the mean($\mu$) and standard deviation ($\sigma$) for the baseline model, as well as the mean MAPE ($\mu_\mathrm{SE}$) and standard deviation ($\sigma_\mathrm{SE}$) for the SE-enhanced version. It also reports the calculated $t$-statistic and $p$-value, along with an indication of whether the difference meets the $p < 0.05$ threshold for statistical significance.

\begin{table}[ht]
\centering
\caption{Paired t-test results comparing WDNN without and with SE blocks over five independent runs. Values are means $\mu$ and $\mu_\mathrm{SE}$, standard deviation $\sigma$ and $\sigma_\mathrm{SE}$ of MAPE (\%) for each target without and with SE blocks, respectively, along with $t$-statistic ($t$-stat), $p$-value, and significance at $p$-value =$0.05$. }
\begin{adjustbox}{max width=\columnwidth}
\begin{tabular}{| l| c| c| c| c| c| c| c|}
\hline
\textbf{Target Variable} & $\mu$ &  $\sigma$  & $\mu_\mathrm{SE}$& $\sigma_\mathrm{SE}$ & \textbf{$t$-stat} & \textbf{$p$-value} & \textbf{Significant?} \\
\hline \hline
Toughness (MPa) & 29.771 & 4.307 & 22.994 & 1.169 & 3.080 & 0.037 & Yes \\ \hline
Strain at Break (\%) & 20.821 & 1.562 & 18.779 & 2.604 & 1.209 & 0.293 & No \\ \hline
Young's Modulus (GPa) & 17.445 & 1.307 & 13.122 & 1.635 & 6.895 & 0.002 & Yes \\ \hline
Yield Strength (MPa) & 19.910 & 2.853 & 15.835 & 0.676 & 3.184 & 0.033 & Yes \\ \hline
Tensile Strength (MPa) & 13.603 & 1.748 & 10.978 & 0.786 & 3.305 & 0.030 & Yes \\ \hline
Print Time (min) & 8.970 & 2.077 & 5.733 & 1.025 & 2.953 & 0.042 & Yes \\ \hline
Weight (g) & 5.312 & 1.507 & 4.178 & 0.765 & 1.799 & 0.146 & No \\ \hline
Material Cost (\$) & 7.556 & 1.128 & 7.035 & 1.191 & 1.082 & 0.340 & No \\ \hline
Overall (MAPE) & 15.422 & 0.527 & 12.334 & 0.547 & 8.292 & 0.001 & Yes \\
\hline
\end{tabular}
\label{tab:paired_ttest_se}
\end{adjustbox}
\end{table}

The results reveal that adding the SE block leads to a notable drop in MAPE for several targets. Significant improvements were observed for Toughness (MPa) ($\mu=29.771$, $\mu_\mathrm{SE}=22.994$, $p=0.037$), Young's Modulus (GPa) ($\mu=17.445$, $\mu_\mathrm{SE}=13.122$, $p=0.002$), Yield Strength (MPa) ($\mu=19.910$, $\mu_\mathrm{SE}=15.835$, $p=0.033$), Tensile Strength (MPa) ($\mu=13.603$, $\mu_\mathrm{SE}=10.978$, $p=0.030$), Print Time (min) ($\mu=8.970$, $\mu_\mathrm{SE}=5.733$, $p=0.042$), and the Overall MAPE score ($\mu=15.422$, $\mu_\mathrm{SE}=12.334$, $p=0.001$). Taken together, these results indicate that SE blocks help the model recalibrate features at the channel level more effectively, which in turn boosts prediction accuracy for both mechanical properties and time-related outputs.

In contrast, the improvements were not statistically significant for strain at break (\%) ($p=0.293$), weight (g) ($p=0.146$), and material cost (\$) ($p=0.340$). This pattern indicates that the SE mechanism may have less impact on these particular targets-perhaps because they are less sensitive to channel attention, or because the baseline WDNN was already capturing the most relevant features for them. Overall, the findings point to the value of applying SE blocks selectively within wide-and-deep architectures, especially when the goal is to improve predictions for mechanically sensitive properties and reduce overall prediction error.

\section{Conclusion}\label{Sec:conclusion}
The current study presented an innovative experimental design guided by LHS and an SE-WDNN for multi-target prediction of economic and mechanical properties in CFRC-AM. From a design space of $4,320$ combinations defined by seven process/material parameters, we used Latin Hypercube Sampling to select 155 representative specimens to maximize broad, space-filling coverage, while the SE-WDNN is learning low- and high-order interactions through channel-wise recalibration. Moreover, the proposed framework addressed two long-standing gaps in previous CFRC-AM work, namely limited parameter coverage and single-target modeling.

In a comprehensive comparative study, the proposed SE-WDNN obtained the best average error (MAPE $ = 12.33\%$) in some benchmark machine learning models, including FFN, CatBoost, RF, XGBoost, KAN. Utilizing a paired t-test across five independent runs (each using different random seeds), we compared SE-WDNN to the same WDNN base models without SE blocks. SE-WDNN resulted in significantly lower per-target errors for Toughness, Young's Modulus, Yield Strength, Tensile Strength, and Print Time, and reduced the model's Overall MAPE\% ($p < 0.05$). Differences for targets such as Strain at Break, Weight, and Material Cost were not significant. From our ablation study, this improvement could be attributed to the SE- block.

Model interpretability via SHAP indicated that the reinforcement strategy was consistently identified as the primary contributory factor for mechanical targets. In particular, toughness, Young's modulus, yield strength, and tensile strength were most affected by the number of reinforced layers, fiber type, and fiber orientation. For Strain at Break, the fiber type was the most important characteristic. On the production side, build time was mainly influenced by the number of reinforced layers and fill density, while weight and material cost were strongly associated with reinforcement characteristics, specifically, the number of reinforced layers and type of continuous fibers. These patterns provide actionable insights for managing mechanical performance and production efficiency considerations in CFRC-AM. 

From an engineering perspective, the SE-WDNN surrogate offers a quick method to explore the CFRC-AM design space within the Mark Two process window. It predicts mechanical and production metrics for possible combinations of the seven manufacturing parameters, which helps reduce reliance on trial-and-error testing. All experiments took place on a Markforged Mark Two using a single tensile-specimen geometry. Therefore, the surrogate is specifically calibrated to this defined workflow. The next step is to validate it on different machines, geometries, and material systems. Additionally, integrating the SE-WDNN framework into optimization or inverse-design routines can allow for the exploration of parameter combinations beyond those tested experimentally.

The results show that LHS-guided sampling and SE-WDNN provide accurate, and multi-target predictions for parameter selection and design exploration in CFRC-AM. While LHS improved coverage, the dataset still covers only a small part of the design space. This limits the ability to capture complex interactions across the full range of parameters. Expanding the sample size and exploring more factor levels could help capture higher-order interactions and improve generalizability. Future work could combine data-rich types of inputs, such as microstructure images or in-situ signals, with tabular features. It could also use active learning or Bayesian optimization loops that sequentially indicate the next most informative experiments to achieve better overall accuracy with a smaller number of experiments. This approach would balance mechanical performance with manufacturing metrics. As a result, it would increase applicability while maintaining the improvements in accuracy and insight that have been shown.

\section*{Acknowledgments}
The authors gratefully acknowledge the support of the Center for Digital and Human-Augmented Manufacturing (CDHAM) and the O’Donnell Data Science and Research Computing Institute at Southern Methodist University for providing resources and infrastructure essential to the completion of this project.

\bibliographystyle{elsarticle-num-names}
\bibliography{6-References}

\appendix
\section{Physical and mechanical properties of Materials}\label{AppA}

\begin{table}[H]
\centering
\caption{Physical and mechanical properties of matrix materials.\cite{MarkforgedInc2025}}\label{tab:plasprops}
\begin{adjustbox}{max width=\columnwidth}
\begin{tabular}{| l c c c|}
\hline
\textbf{Property} & \textbf{Test (ASTM)} & \textbf{Nylon} & \textbf{Onyx} \\
\hline \hline
Tensile Stress at Yield (MPa)     & D638        & $40$   & 51   \\ \hline
Tensile Modulus (GPa)             & D638        & 1.7  & 2.4  \\ \hline
Tensile Stress at Break (MPa)     & D638        & 36   & 37   \\ \hline
Tensile Strain at Break (\%)      & D638        & 150  & 25   \\ \hline
Flexural Strength (MPa)           & D790        & 50   & 71   \\ \hline
Flexural Modulus (GPa)            & D790        & 1.4  & 3.0  \\ \hline
Izod Impact - Notched (J/m)       & D256-10 A   & 110  & 330  \\ \hline
Density (g/cm$^3$)                & --          & 1.1  & 1.2  \\
\hline
\end{tabular}
\end{adjustbox}
\end{table}
\begin{table}[H]
\centering
\centering
\caption{Physical and mechanical properties of continuous fibers used for reinforcement.\cite{MarkforgedInc2025}}\label{tab:fiberprop}
\begin{adjustbox}{max width=\columnwidth}
\begin{tabular}{| l c c c c|}
\hline
\textbf{Property} & \textbf{Test (ASTM)} & \textbf{Carbon} & \textbf{Fiberglass} & \textbf{Kevlar} \\
\hline \hline
Tensile Strength (MPa)            & D3039       & 800  & 590  & 610  \\ \hline
Tensile Modulus (GPa)              & D3039       & 60   & 21   & 27   \\ \hline
Tensile Strain at Break (\%)       & D3039       & 1.5  & 3.8  & 2.7  \\ \hline
Flexural Strength (MPa)            & D790        & 540  & 200  & 240  \\ \hline
Flexural Modulus (GPa)              & D790        & 51   & 22   & 26   \\ \hline
Flexural Strain at Break (\%)       & D790        & 1.2  & 1.1  & 2.1  \\ \hline
Compressive Strength (MPa)         & D6641       & 420  & 180  & 130  \\ \hline
Compressive Modulus (GPa)          & D6641       & 62   & 24   & 25   \\ \hline
Compressive Strain at Break (\%)   & D6641       & 0.7  & --   & 1.5  \\ \hline
Izod Impact - Notched (J/m)        & D256-10 A   & 960  & 2600 & 2000 \\ \hline
Density (g/cm$^3$)                  & --          & 1.4  & 1.5  & 1.2  \\
\hline
\end{tabular}
\end{adjustbox}
\end{table}
\end{document}